\documentclass[twocolumn,floatfix,aps]{revtex4}
\usepackage{bm}
\usepackage{graphicx,color}

\usepackage{amsmath,amssymb}

\begin{document}

\title{Contrarian compulsions produce exotic time dependent flocking of active particles} 
\author{L L Bonilla$^{1,2,*}$ and C. Trenado$^1$}
\affiliation{G. Mill\'an Institute for Fluid Dynamics, Nanoscience and
Industrial Mathematics, and Department of Materials Science and Engineering and Chemical Engineering, Universidad Carlos III de Madrid, 28911 Legan\'es,
Spain\\
$^2$Courant Institute for Mathematical Sciences, New York University, 251 Mercer St., New York, NY 10012\\
$^*$Corresponding author. E-mail: bonilla@ing.uc3m.es} 
\date{\today}
\begin{abstract}
Animals having a trend to align their velocities to an average of those of their neighbors may flock as illustrated by the Vicsek model and its variants. If, in addition, they feel a systematic contrarian trend, the result may be a time periodic adjustment of the flock or period doubling in time. These exotic new phases are predicted from kinetic theory and numerically found in a modified two-dimensional Vicsek model of self-propelled particles. Numerical simulations demonstrate striking effects of alignment noise on the polarization order parameter measuring particle flocking: maximum polarization length is achieved at an optimal nonzero noise level. When contrarian compulsions are more likely than conformist ones, non-uniform polarized phases appear as the noise surpasses threshold.
\end{abstract}


\maketitle

\section{Introduction} \label{sec:1}
The ability to convert free energy into systematic motion characterizes active matter \cite{ram10,vic12,mar13,bi16,hak17}. Thus collective behavior of active particles is a central aspect of the subject. Active particles may be cells in confluent motion \cite{hak17,mal17}, bacteria \cite{igo01,igo04,bal15}, spermatozoa \cite{cre16}, animals \cite{cou05,ton95,ton05,bal08,att14,jia17,cav18,mig18}, active gels \cite{sim02,jul07}, colloids \cite{the12}, active-liquid metamaterials \cite{sou17}, or interacting robots \cite{rub14}. Patterns, stability and morphology of swarms depend on size and have been analyzed in Ref.~\cite{ors06} for a class of models. 

Collective motion of active particles is often modeled by the Vicsek model (VM) \cite{vic95} or its variants \cite{ram10,vic12,mar13,mig18,ihl11,cho12,ihl15,ihl16}. Typically, the velocities of $N$ particles moving with equal speed are updated so that the velocity of each particle adopts the direction of the average velocity of its close neighbors with some alignment noise ({\em conformist or majority rule}) \cite{vic95,vic12}. These systems exhibit a phase transition from disordered to coherent behavior that is an example of spontaneous symmetry breaking out of equilibrium. When the particles are inside a box below a critical size, there is a continuous  bifurcation of the VM with forward update: the system bifurcates from a disordered state with uniform density to an ordered state characterized by nonzero average speed of the particles \cite{vic95}. For box size larger than critical, bifurcation is discontinuous and a variety of patterns are possible \cite{gre04,cha08,sol15,hue08}. Independently of the box size, the bifurcation seems to be continuous for VM with backward update \cite{bag09}. 

There are examples of more complex individual behavior in active particles. For example, instead of moving linearly, {\em E. coli} bacteria swim clockwise in circular trajectories near walls \cite{dil05,lau06}. Simpler artificial microswimmers can also be induced to move circularly about solid surfaces \cite{tak14} and autonomous motion of active colloids is reviewed in Ref.~\cite{ebb16}. Recently, Liebchen and Levis have proposed a model of chiral active matter to explain this motion \cite{lie17}. In two dimensions, each particle $i$ moves with constant speed $v$ and is parallel to its polarization vector. The angular velocity of the latter is the sum of three terms: a constant rotation, a white noise, and a Kuramoto coupling between the angles of particles inside the circle of influence of the particle $i$ \cite{lie17}. This model exhibits a flocking transition to one or several clusters comprising particles rotating in synchrony (microflocking) and a variety of patterns. 

In this paper, we explore a different mechanism to attain synchronous rotation in small clusters. We consider a two-dimensional (2D) VM with forward update. Active particles may be {\em conformist} and align their velocities to the average velocity of their neighbors with probability $1-p$, or be {\em contrarian} and move opposite to the average angle with probability $p\in[0,1]$. This choice makes the VM similar to a Kuramoto model of phase synchronization \cite{kur75,ace05} with conformist and contrarian oscillators \cite{hon11}. It is different from the {\em self-centered conformist rule} of Ref.~\cite{rom14}. There, the particles inside the circle of influence of a given one are considered its neighbors only if their velocities are within a certain angular sector about the velocity of the particle. Particles with velocities having directions outside the sector are ignored, even if they are close to the considered particle. In this case, the flocking transition becomes discontinuous for sufficiently narrow sectors, but flocking is still stationary \cite{rom14}. In opinion formation models \cite{heg02,kur16}, our VM may mimic the conflict between reaching consensus with others and keeping a contrarian opinion. Experiments on imitative behavior in emergency escape of human crowds have shown that sometimes avoiding the majority is the best survival strategy \cite{hag17}. Our VM is a step towards exploring this behavior.

Besides disordered and ordered phases with almost zero and finite stationary polarization, respectively, numerical simulations show that our VM exhibits novel exotic phases (stable solutions) with time periodic polarization order parameter. In the ordered phases, flocking is  heterogeneous in space and reminiscent of microflocking in Ref.~\cite{lie17}. Strikingly, in the presence of contrarian compulsions, increasing the Vicsek alignment noise may favor order in two ways. For small $p$, there is a nonzero optimal noise value for which polarization is maximal and reflects a trend to homogeneous flocking. When contrarian compulsions are prevalent ($p$ closer to 1), increasing the alignment noise may transform incoherent particle motion to a phase displaying time periodic polarization with period 2. If we relax the contrarian rule so that particles may select deflection by some fixed large angle $\xi_0$ ($\pi/2<\xi_0<\pi$) measured counterclockwise from the average direction, the flocking order parameter may oscillate periodically in time (with period different from 2, which occurs for $\xi_0=\pi$). Active particles perform rotations or oscillations besides the collective translation characterizing the ordered phase of the standard VM. Rotation allows active particles to explore larger regions of space and may be advantageous in emergency escape of a crowd from a confined region with several exits \cite{hag17}. 

To interpret and understand the results of our numerical simulations, it is convenient to analyze near the transition to flocking a kinetic theory proposed by T. Ihle for the standard VM \cite{ihl11}. The exotic phases appear as Hopf and period doubling bifurcations from the uniform distribution function at a critical value of the noise. Our bifurcation calculations follow those we have developed for the standard VM \cite{BT18}. We find that the equation for the complex amplitude of the bifurcating solution is a modified complex Ginzburg-Landau equation (CGLE). This equation has a uniform rotating wave solution whose polarization has critical exponent $1/2$ at the bifurcation point. Deviation of the numerically obtained polarization from this theoretical curve indicates heterogeneous flocking.

The rest of the paper is as follows. The modified VM and its formulation in terms of a kinetic equation for the distribution function are described in Section \ref{sec:2}. Section \ref{sec:3} analyzes the linear stability of the uniform distribution function corresponding to the disordered state of the particles. Depending on the weights of conformist and contrarian or almost contrarian rules, the ordered states appear as period-doubling, Hopf or pitchfork bifurcations. However, conservation of the number of particles implies that the amplitude equations for the complex amplitudes of the corresponding bifurcating modes are coupled with an equation for the particle density. For the standard VM, the pitchfork bifurcation is studied in Ref.~\cite{BT18}. Section \ref{sec:4} describes the general use of the Chapman-Enskog method of Ref.~\cite{BT18} for other bifurcations. Section \ref{sec:5} discusses the solutions of the amplitude (bifurcation) equations in the cases of Hopf and period doubling bifurcations. The case of the pitchfork bifurcation is presented in  Section \ref{sec:6}. The results of numerical simulations of the VM with contrarian and almost contrarian compulsions are presented and explained using the kinetic equation in Section \ref{sec:7}. Finally, Section \ref{sec:8} contains our final remarks and discussion of our results whereas the Appendices describe our nondimensionalization of the Vicsek model and different technical matters.

\section{Model and kinetic equation} \label{sec:2}
\subsection{Model}
Active particles placed in a square box of side $L$ with periodic boundary conditions tend to align their velocities to an average of those of their neighbors:
\begin{eqnarray}
\theta_j(t+1)=\mbox{Arg}\!\left(\sum_{|\mathbf{x}_l-\mathbf{x}_j|<R_0} e^{i\theta_l(t)}\right)\!+\xi_j(t), \label{eq1}\\
\mathbf{x}_j(t+1)=\mathbf{x}_j(t)+(\cos\theta_j(t+1),\sin\theta_j(t+1)).\label{eq2}
\end{eqnarray}
Here $\mathbf{x}_j(t)$ and $\theta_j(t)$ are position and phase of the $j$th particle velocity at time $t$, $j=1,\ldots, N$.  In Eq.~\eqref{eq1}, neighbors are all particles inside a circle of radius $R_0$ centered at particle $j$ \cite{vic95}. This {\em metric} concept of neighborhood seems to describe appropriately insect swarms \cite{att14}. Nondimensional time step and particle speed are one. See Appendix \ref{ap:1} for the relation to dimensional units. $\xi_j(t)$ are independent identically distributed (i.i.d.) random alignment noises, selected with probability density 
\begin{equation}\label{eq3}
g(\xi)= p\,\delta(\xi-\xi_0) + \frac{1-p}{\eta}\chi_{[-\eta/2,\eta/2]}(\xi), \quad 0\leq p\leq 1.
\end{equation}
Here $\xi_0\in(-\pi/2,\pi]$, and $\chi_{[-\eta/2,\eta/2]}(\xi)$ is 1 for $-\eta/2<\xi<\eta/2$ and 0 otherwise. The parameter $\eta$ measures the width of the alignment noise and can we thought of as a tolerance to failure in the alignment rule. If $\xi_0=\pi$, particles experience contrarian compulsions with probability $p$, and conformist compulsions toward the average angle of their neighbors with probability $(1-p)$. If $-\pi/2<\xi_0<\pi$, particles experience a large deflection instead of a perfectly contrarian alignment. We speak of {\em almost contrarian compulsions} because the ordered phases are similar rotating wave phases for all angles of deflection in $(-\pi/2,\pi)$. The numerical method used to implement Eq.~\eqref{eq3} is described in Appendix \ref{ap:2}. For $p=0$, Eqs.~\eqref{eq1}-\eqref{eq3} are the standard VM (with forward update). 

Collective consensus is quantified by the complex order parameter 
\begin{eqnarray}
Z=W\, e^{i\Upsilon}=\frac{1}{N}\sum_{j=1}^N e^{i\theta_j}, \label{eq4}
\end{eqnarray}
whose amplitude $0<W<1$ (polarization) measures macroscopic coherence of the particles and $\Upsilon$ is their average phase. For the standard VM ($p=0$), increasing the average number of particles inside the region of influence, $M=N\pi R_0^2/L^2$, favors flocking as more and more particles try to move together. Instead, increasing the alignment noise $\eta$ tends to destroy flocks, as it dilutes the efficacy of the alignment rule. Then there are a critical value of $M$ above which the polarization is $W>0$ and below which $W=0$ (in the limit as $N\to\infty$). The alignment noise also has a critical value, but now it has $W>0$ below and $W=0$ above threshold. For our model, these two features hold for $p$ small enough with the difference that the ordered phases are not necessarily time independent. For $p$ closer to 1, increasing $M$ also favors flocking because more and more particles interact and get to move together. However, we shall see that the alignment noise $\eta$ may favor flocking because it dilutes the strength of the conformist rule and it allows the contrarian or almost contrarian rule to form clusters of particles moving synchronously. We could replace a similar density $\chi_{[-\tilde{\eta}/2,\tilde{\eta}/2]}(\xi-\xi_0)/\tilde{\eta}$ instead of $\delta(\xi-\xi_0)$ in Eq.~\eqref{eq3}, thus adding a more realistic tolerance $\tilde{\eta}$ to the choice of deflection angle $\xi_0$. However, the results would be qualitatively similar to those reported here.

\subsection{Kinetic equation}
In the limit as the number of particles goes to infinity, it is possible to derive a kinetic equation for the VM following Refs.~\cite{ihl11,ihl16}. The $N$-particle probability density, $P_N$ satisfies the following exact equation:
\begin{widetext}
\begin{eqnarray}
P(\theta^{(N)}\!,\mathbf{x}^{(N)}\!+\mathbf{v}^{(N)}\!,t+1)=\!\int_{[-\pi,\pi]^N}\!\int_{[-\pi,\pi]^N}\! P(\tilde{\theta}^{(N)}\!,\mathbf{x}^{(N)}\!,t)\left(\prod_{i=1}^N\hat{\delta}(\theta_i-\xi_i-\Phi_i(\tilde{\theta}^{(N)}\!,\mathbf{x}^{(N)}\!,t))\, d\xi_id\theta_i\right)\!.
\label{eq5}
\end{eqnarray}
\end{widetext}
Here $\theta^{(N)}=\theta_1,\ldots,\theta_N$, with similar expressions for $\tilde{\theta}^{(N)}$, $\xi^{(N)}$, $\mathbf{x}^{(N)}$, $\mathbf{v}^{(N)}$. All integrations over angles and noises go from $-\pi$ to $\pi$, and $\Phi_j\!=$Arg$(\sum_{|\mathbf{x}_l-\mathbf{x}_j|<R_0}e^{i\tilde{\theta}_l})$ is the average direction of the vector sum of all particle velocities (including particle $j$) inside the interaction circle of radius $R_0$ centered at $\mathbf{x}_j$. $\hat{\delta}(x)=\sum_{l=-\infty}^\infty\delta(x+2\pi l)$ is a periodized delta function that incorporates the ``collision rule'' Eq.~\eqref{eq1}, and the integrals over $\xi_j$ average over the noises. 

From the exact equation \eqref{eq5}, we derive an approximate equation for the distribution function $f$ such that $f(\theta,\textbf{x},t)\, d\textbf{x}\, d\theta$ is the number of particles in an area $d\textbf{x}$ centered at position $\mathbf{x}$ that move into a direction between $\theta$ and $\theta+d\theta$ at time $t$. Let us for a moment recall models that are somewhat more complicated than the VM \cite{hub04,hak17}. In these models, particles have a repelling core $r^*$ related to the size of the animals we are modeling, an influence zone of radius $R_0$ ($R_0>r^*$) and, possibly, a larger attraction zone \cite{hub04,sep13}. On nondimensional times of order $r^*$ (recall that the speed is one in our units), some particles interact and there are strong particle-particle correlations. On much longer times, $t\gg r^*$, memory of these strong interactions is erased and we can assume that all particles are independent and identically distributed before undergoing a collision given by Eq.~\eqref{eq1}. Then the $N$-particle probability density is product of $N$ one-particle probability densities $f(\mathbf{x}_j,\theta_j,t)/N$, $j=1,\ldots, N$, ($f(\mathbf{x}_j,\theta_j,t)$ is the one-particle distribution function):
\begin{equation}
P_N(\theta^{(N)},\mathbf{x}^{(N)},t)=\prod_{i=1}^N\frac{f(\mathbf{x}_j,\theta_j,t)}{N}. \label{eq6}
\end{equation}
This is the {\em molecular chaos assumption} used to derive the Boltzmann equation \cite{hua87}. Eq.~\eqref{eq6} is appropriate for a time discrete model with unit time step if $r^*\ll 1$ in our nondimensional units. For the VM, $r^*=0$, and the molecular chaos assumption is very reasonable. See, however, Ref.~\cite{cho15} for a discussion of corrections to molecular chaos in the limit $R_0\ll 1$. Eq.~\eqref{eq6} produces the formula
\begin{widetext}
\begin{eqnarray}
\int_{[-\pi,\pi]^N}\!\int_{[0,L]^{2N}} \sum_{j=1}^N\delta(\theta-\theta_j)\delta(\mathbf{x}-\mathbf{x}_j)\, P_N(\theta^{(N)},\mathbf{x}^{(N)},t)\prod_{i=1}^N(d\theta_id\mathbf{x}_i) 
 = f(\theta,\mathbf{x},t), \label{eq7}
\end{eqnarray}
\end{widetext}
where the $\mathbf{x}_j$ are integrated on the box $[0,L]\times[0,L]$. 

Now let us multiply the equation resulting from Eqs.{\eqref{eq5} and \eqref{eq6} by $\sum_{j=1}^N\delta(\theta-\theta_j)\delta(\mathbf{x}-\mathbf{x}_j)$ and integrate over all angles and positions to obtain an equation for the distribution function. The result is \cite{ihl16}
\begin{eqnarray}
f(\textbf{x}+\textbf{v},\theta,t+1)=C[f](\theta,\textbf{x},t),\quad\textbf{v}=(\cos\theta,\sin\theta), \label{eq8}
\end{eqnarray}
\begin{widetext}
\begin{eqnarray}
&&C[f]\!=\!\int^{\pi}_{-\pi}\! d\xi \, g(\xi)\sum_{n=1}^N\left(\begin{array}{c} N-1\\ n-1 \\ \end{array}\right)
\int_{[-\pi,\pi]^{n}} \hat{\delta}(\theta-\xi-\Phi_1(\tilde{\theta}_1,\ldots,\tilde{\theta}_n))\, f(\textbf{x},\tilde{\theta}_1,t)\left(1-\frac{M_R(\mathbf{x},t)}{N}\right)^{N-n}\nonumber\\
&&\quad\quad\times \prod_{i=2}^{n}\!\left[\int_{|\mathbf{x}_i-\mathbf{x}|<R_0}\frac{f(\textbf{x}_i,\tilde{\theta}_i,t)}{N} d\textbf{x}_id\tilde{\theta}_i\right]\! d\tilde{\theta}_1, \quad\quad\Phi_1(\tilde{\theta}_1,\ldots,\tilde{\theta}_n)=\mbox{Arg}\left(\sum_{j=1}^n e^{i\tilde{\theta}_j}\right)\!, \label{eq9}\\
 &&M_R(\textbf{x},t)= \int_{|\mathbf{x}'-\mathbf{x}|<R_0}\rho(\textbf{x}',t)\, d\mathbf{x}',\quad \rho(\mathbf{x},t)=\int_{-\pi}^\pi f(\mathbf{x},\theta,t)\, d\theta,\quad\int\rho(\mathbf{x},t) d\mathbf{x}=N.\label{eq10}
\end{eqnarray}
\end{widetext}
 In Eq.~\eqref{eq9}, $n$ is the number of particles inside the interaction circle of radius $R_0$ about particle 1 (the latter included). The average number of particles inside an interaction circle about position $\mathbf{x}$ is $M_R(\textbf{x},t)$, given by Eq.~\eqref{eq10}. The combinatorial factor in Eq.~\eqref{eq9} counts the number of possible selections of neighbors of particle 1 (excluding the latter) out of the $N-1$ other particles. The factor $(1-M_R/N)^{N-n}$ in Eq.~\eqref{eq9} gives the probability that the particles $n+1,\ldots,N$ are not within interaction distance of particle 1. The factor $\prod_{i=1}^n\int_{|\mathbf{x}_i-\mathbf{x}|<R_0}f(\textbf{x}_i,\theta_i,t) d\mathbf{x}_i$ is the probability that particles $2,\ldots, n$ be within interaction distance of particle 1 times their angular distribution, given that they are within the interaction distance. In Eq.~\eqref{eq9}, $M_R(\textbf{x},t)$ given by Eq.~\eqref{eq10}, is the average number of particles inside the circle of influence about $\mathbf{x}$. When we integrate Eq. \eqref{eq9} over $\theta$, we find that the particle density immediately after collisions equals that before:
\begin{equation}
\int C[f](\theta,\textbf{x},t)\, d\theta=\rho(\mathbf{x},t). \label{eq11}
\end{equation}

In this paper, we shall assume that the average density $\rho_0=N/L^2$ is large, although the resulting approximations are reasonably accurate for $\rho_0$ larger than one. Then, for $n/N\ll 1$ as $N\to\infty$, the combinatorial factor times $(1-M_R/N)^{N-n}$ becomes
\begin{equation*}
\frac{(N-1)!}{(n-1)!(N-n)!}\left(1-\frac{M_R}{N}\right)^{N-n}\sim \frac{N^{n-1}}{(n-1)!}\, e^{-M_R},
\end{equation*}
and Eq.\eqref{eq9} produces an Enskog-type collision operator \cite{ihl16}
\begin{widetext}
\begin{eqnarray}
C_E[f]\!=\!\int^{\pi}_{-\pi}\! d\xi \, g(\xi) e^{-M_R(\mathbf{x},t)}\!\sum_{n=1}^{\infty}\int_{[-\pi,\pi]^{n}}\!\!\frac{\hat{\delta}(\theta-\xi-\Phi_1(\theta_1,\ldots,\theta_n))}{(n-1)!}
 f(\textbf{x},\theta_1,t)\!\prod_{i=2}^{n}\!\left[\int_{|\mathbf{x}_i-\mathbf{x}|<R_0}f(\textbf{x}_i,\theta_i,t) d\textbf{x}_id\theta_i\right]\!d\theta_1, \label{eq12}
 \end{eqnarray}
 \end{widetext}
which we shall use henceforth. In the integrals of Eq.~\eqref{eq12}, we have used $\theta_j$ instead of the $\tilde{\theta}_j$ of Eq.~\eqref{eq9}. For active particles in a disordered state, the density $\rho(\mathbf{x},t)$ equals the constant average density, $\rho_0=N/L^2$, and the uniform distribution function, $f_0=\rho_0/(2\pi)$, is a fixed point of the collision operators: 
\begin{eqnarray}
C[f_0]=f_0,\quad C_E[f_0]=f_0. \label{eq13}
\end{eqnarray}

\section{Linear stability} \label{sec:3}
\subsection{Eigenvalue problem}
To study the linear stability of the uniform distribution function, we insert $f=f_0+\tilde{\epsilon}\, \tilde{f}(\theta,\textbf{x},t)$, $\tilde{\epsilon}\ll 1$, and ignore quadratic terms, thereby obtaining
\begin{widetext}
\begin{eqnarray}
\tilde{f}(\theta,\textbf{x}+\mathbf{v},t+1)\!=\!\sum_{n=1}^\infty\!\frac{e^{-M}\!\left(\frac{M}{2\pi}\right)^{n-1} }{(n-1)!}\!\! \int^{\pi}_{-\pi}\! d\xi\, g(\xi)\!\int_{[-\pi,\pi]^n}\!\hat{\delta}(\theta-\xi-\Phi_1)\!\left[\tilde{f}(\mathbf{x},\theta_1,t)+\frac{n-1}{\pi R_0^2}\! \int_{|\mathbf{x}'-\mathbf{x}|<R_0}\!\tilde{f}(\mathbf{x}',\theta_1,t)d\mathbf{x}' \right]\!  \nonumber\\
\times\prod_{i=1}^nd\theta_i -\frac{M}{2\pi}\frac{1}{\pi R_0^2}\int_{|\mathbf{x}'-\mathbf{x}|<R_0}\int^{\pi}_{-\pi}\tilde{f}(\theta',\textbf{x}',t) d\textbf{x}' d\theta'.\quad \label{eq14}
\end{eqnarray}
\end{widetext}
We now separate variables by inserting $\tilde{f}=h(t)\, g(\mathbf{x},\theta)$ in Eq.~\eqref{eq14} (divided by $\tilde{f}$). Then $h(t+1)/h(t)$ equals a function of $\mathbf{x}$ and $\theta$. Thus both sides equal a constant $Q$, from which $h(t)=Q^t$. Since $\tilde{f}$ satisfies periodic boundary conditions on the box of side $L$, it is a Fourier series of plane waves $e^{i\mathbf{K}\cdot\mathbf{x}}$, where $\mathbf{K}=2\pi (n,m)/L$, with integer values of $n$ and $m$. Then we can set $\tilde{f}=Q^t e^{i\mathbf{K}\cdot\mathbf{x}}\varphi(\theta)$ in Eq.~\eqref{eq14} and obtain the eigenvalue problem of finding $Q$ for which there is a nonzero solution of
\begin{eqnarray}
Qe^{i\mathbf{K}\cdot\mathbf{v}}\varphi- C^{(1)}[\varphi]=0,\label{eq15}
\end{eqnarray}
\begin{widetext}
\begin{eqnarray}
\!\!\!C^{(1)}[\varphi]\!=\!\frac{2J_1(|\mathbf{K}|R_0)}{|\mathbf{K}|R_0}\left[\sum_{n=1}^\infty\frac{e^{-M}}{(n-1)!}\left(\frac{M}{2\pi}\right)^{n-1}\!\! \left(n-1+\frac{|\mathbf{K}|R_0}{2J_1(|\mathbf{K}|R_0)}\right)\!\!\int^{\pi}_{-\pi}\! d\xi\, g(\xi)\!
\!\int_{[-\pi,\pi]^n}\hat{\delta}(\theta-\xi-\Phi_1)\right.\nonumber\\
\times\varphi(\theta_1)\prod_{i=1}^nd\theta_i\left. -\frac{M}{2\pi}\int^{\pi}_{-\pi}\varphi(\theta') d\theta'\right]\!.  \label{eq16}
\end{eqnarray}
\end{widetext}
Here $M=\rho_0\pi R_0^2$. We have $C^{(1)}[1]=1$, and therefore any function independent of the angle $\theta$ (for example, the uniform distribution $f_0=\rho_0/(2\pi)$) solves Eq.~\eqref{eq15} with  $|\mathbf{K}|=0$ and $Q=1$. 

We now insert the Fourier expansion $\varphi(\theta)=\sum_{j=-\infty}^\infty\varphi_j e^{ij\theta}$ in Eqs.~\eqref{eq15}-\eqref{eq16} and use that $C^{(1)}[e^{il\theta}]_j=0$ for $j\neq l$ \cite{ihl16} to obtain
\begin{eqnarray}
\sum_{j=-\infty}^\infty \{C^{(1)}[e^{ij\theta}]_j\delta_{jl}-Q(e^{i\mathbf{K}\cdot(\cos\theta,\sin\theta)+ij\theta})_l\} \varphi_{j}=0. \label{eq17}
\end{eqnarray}
Then $Q$ are the zeros of the eigenvalues $\lambda(Q)$ of the matrix $\mathcal{M}_{jl}(Q)$ (cf. Appendix \ref{ap:2}): 
\begin{widetext}
\begin{eqnarray}
&&\mathcal{M}_{lj}(Q)=Q\, (e^{i\mathbf{K}\cdot(\cos\theta,\sin\theta)+ij\theta})_l - C^{(1)}[e^{ij\theta}]_j \delta_{jl}, \quad (e^{i\mathbf{K}\cdot(\cos\theta,\sin\theta)+ij\theta})_l =e^{i(j-l)[\mbox{\footnotesize{Arg}}(\mathbf{K})+\pi/2]} J_{j-l}(|\mathbf{K}|), \nonumber\\
&&C^{(1)}[e^{ij\theta}]_j=\!\left(\int^{\pi}_{-\pi}\! e^{-ij\xi} g(\xi)\,d\xi\right)\!\sum_{n=1}^\infty\frac{M^{n-1}e^{-M}}{(n-1)!}\!\!\left[(n-1)\frac{2J_1(|\mathbf{K}|R_0)}{|\mathbf{K}|R_0}+1\right]\!\int_{[-\pi,\pi]^n} e^{ij(\theta_1-\Phi_1)}\prod_{l=1}^n \frac{d\theta_l}{2\pi} .  \label{eq18}
\end{eqnarray}
\end{widetext}
We have not succeeded in finding the eigenvalues of the matrix $\mathcal{M}_{jl}(Q)$ for arbitrary $\mathbf{K}$. However, for small $|\mathbf{K}|$, the off-diagonal elements of the matrix are small compared to its diagonal elements, $Q\, J_0(|\mathbf{K}|)-C^{(1)}[e^{ij\theta}]_j$. Then we can use second-order perturbation theory to derive the following formula (cf. Appendix \ref{ap:2}):
\begin{eqnarray}
Q_1(|\mathbf{K}|,\eta)\!\approx\!\frac{C^{(1)}[e^{i\theta}]_1}{J_0(|\mathbf{K}|)}\!\!\left[1\!-\!\frac{[J_1(|\mathbf{K}|)/J_0(|\mathbf{K}|)]^2}{1- C^{(1)}[e^{i2\theta}]_2/C^{(1)}[e^{i\theta}]_1\!}\right]\!,  \label{eq19}
\end{eqnarray}
in which $C^{(1)}[e^{ij\theta}]_j$ are given by Eq.{\eqref{eq18}. 

The uniform distribution becomes unstable when one eigenvalue $Q$ moves outside the unit circle in the complex plane. It turns out that $Q_1$ has the largest modulus for small $|\mathbf{K}|$, as we argue below. Depending on the parameters $M$ and $\rho_0$, the equation $|Q_1(|\mathbf{K}|,\eta)|=1$ may have different solution branches $\eta(|\mathbf{K}|)$ for $|\mathbf{K}|>1$, \cite{BT18}. However, for these branches, the off-diagonal elements of the matrix $\mathcal{M}_{jl}(Q)$ are no longer small compared to its diagonal elements, the regular perturbation theory is no longer valid, and we ignore them.

\begin{figure}[h]
\begin{center}
\includegraphics[width=8cm]{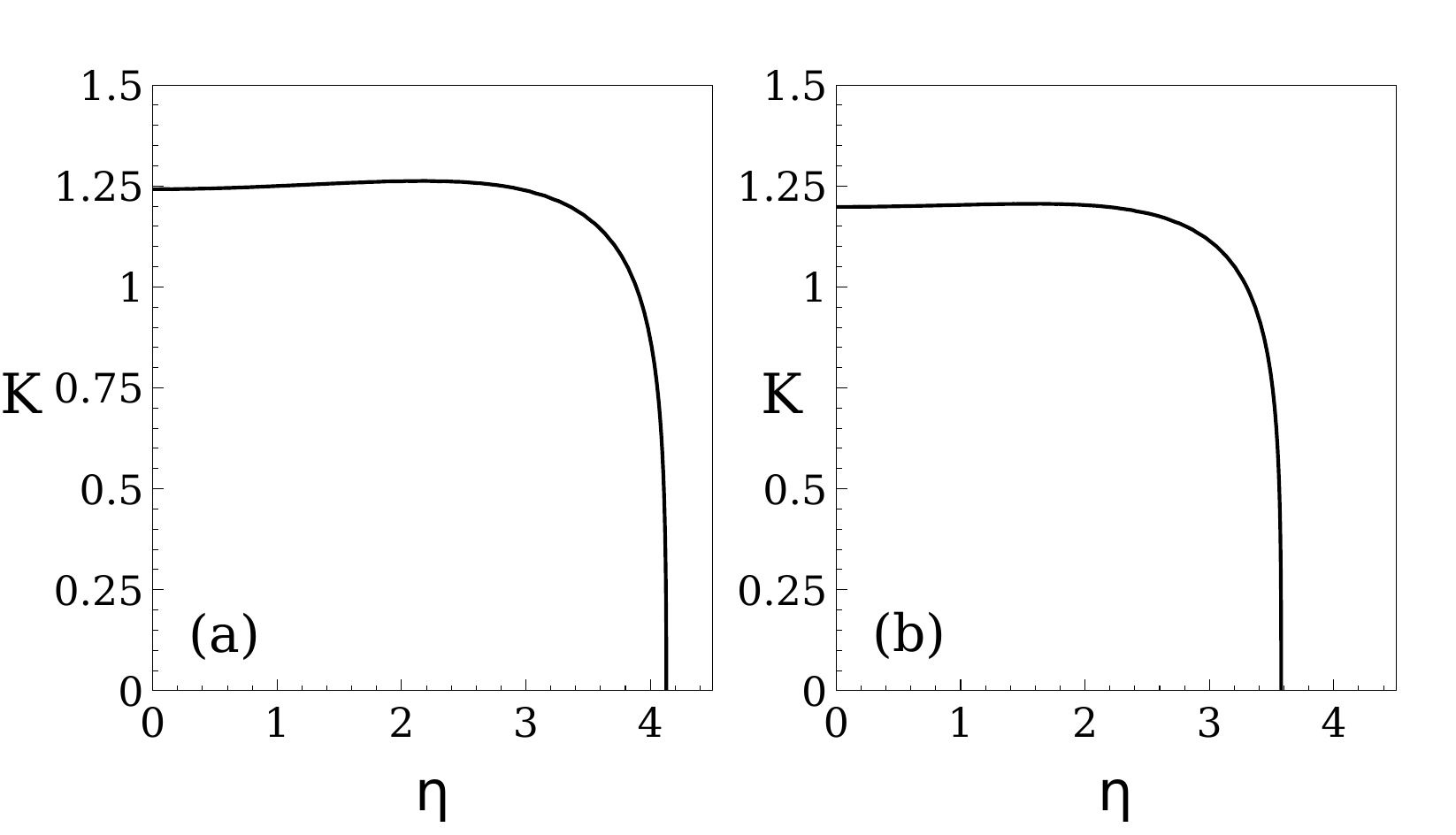}
\end{center}
\caption{Critical wave vector $K=|\mathbf{K}|$ versus $\eta$ for $M=7$ and (a) $p=0$, $\rho_0=5$; (b) $p=0.1$, $\xi_0=3\pi/4$, $\rho_0=10$. These curves are obtained by solving $|Q_1(K,\eta)|=1$, where $Q_1(K,\eta)$ is given by Eq.~\eqref{eq19}. Finite box size $L$ means that the wave vectors in this figure may only take on the values $\mathbf{K}=2\pi(n,m)/L$ with integer $n$ and $m$.\label{fig1}}
\end{figure}

Figs.~\ref{fig1}(a) and (b) show the solution curves of $|Q_1(|\mathbf{K}|,\eta)|=1$ for $p=0$, $\rho_0=5$ (standard VM) and for $p=0.1$, $\xi_0=3\pi/4$ and $\rho_0=10$, respectively. Similar curves are found for the contrarian case, $\xi_0=\pi$. In all cases, the uniform distribution becomes first unstable for the noise corresponding to zero wave number. Within our approximations, this justifies that the largest value of the multiplier $Q_1$ is attained at zero wave number. 

For $|\mathbf{K}|=0$, in the limit as $M\gg 1$, we have the eigenvalues \cite{BT18}
\begin{eqnarray}
Q_1\!&\sim& \frac{\sqrt{\pi M}}{2}\int^{\pi}_{-\pi}\! e^{-i\xi} g(\xi)d\xi\nonumber\\
&=&\frac{\sqrt{\pi M}}{2}\!\left(p\, e^{-i\xi_0}+\frac{2(1-p)}{\eta}\sin\frac{\eta}{2}\right)\!,  \label{eq20}\\
Q_2\!&\sim&\frac{1}{2}\int_{-\pi}^\pi e^{-i2\xi}g(\xi)d\xi,\label{eq21}\\
Q_3\!&\sim&-\frac{3\ln\pi}{2\pi}\int_{-\pi}^\pi e^{-i3\xi}g(\xi)d\xi,\label{eq22}\\
Q_{4}\!&\sim&\frac{1}{12M}\int_{-\pi}^\pi e^{-i4\xi}g(\xi)d\xi, \label{eq23}
\end{eqnarray}
and so on, with eigenfunctions $\varphi_j(\theta)=e^{ij\theta}$ (cf. Ref.~\cite{ihl16} for $Q_1$ and $Q_2$ and the standard VM). As $|\int_{-\pi}^\pi e^{-ij\xi}g(\xi)d\xi|\leq\int_{-\pi}^\pi g(\xi)d\xi=1$, $|Q_j|\leq 1$ for $j>1$ in the limit as $M\gg 1$. For $|\mathbf{K}|=0$, the eigenvalue with largest modulus is therefore $Q_1$, which is the only one that can exit the unit circle in the complex plane.

\begin{figure}[h]
\begin{center}
\includegraphics[width=8.5cm]{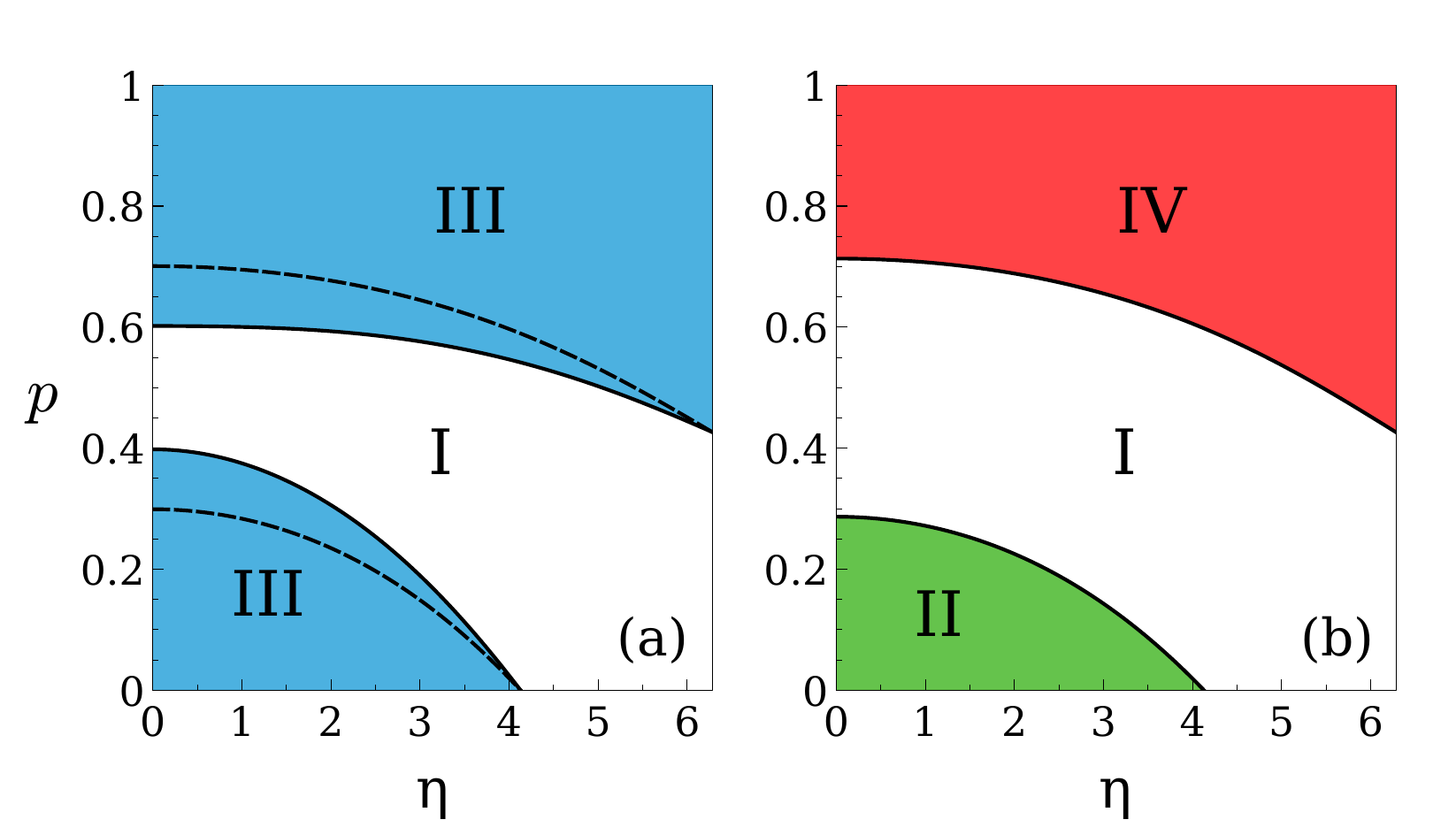}
\end{center}
\caption{Phase diagram of $p$ versus $\eta$ obtained from kinetic theory. Stable phases at each region are: I (incoherent), II (stationary), III (rotating wave), IV (period 2). Parameter values are $M=\pi R_0^2N/L^2=7$, and: (a) $\xi_0=3\pi/4$ (solid line) and $\xi=9\pi/10$ (dashed line); (b) $\xi_0=\pi$. For $p=0$, the lines separating Regions I and III and I and II in the lower half of these panels intersect at the same point, $\eta_c\approx 4.13$, which is the critical noise of the standard VM. Increasing the noise decreases order when $\eta$ crosses the lower critical lines and increases it when $\eta$ crosses the upper critical lines. There is a narrow interval of $p$ values where phase I is stable for any $\eta$. This interval disappears for large enough $M$. \label{fig2}}
\end{figure}

\begin{figure}[h]
\begin{center}
\includegraphics[width=8cm]{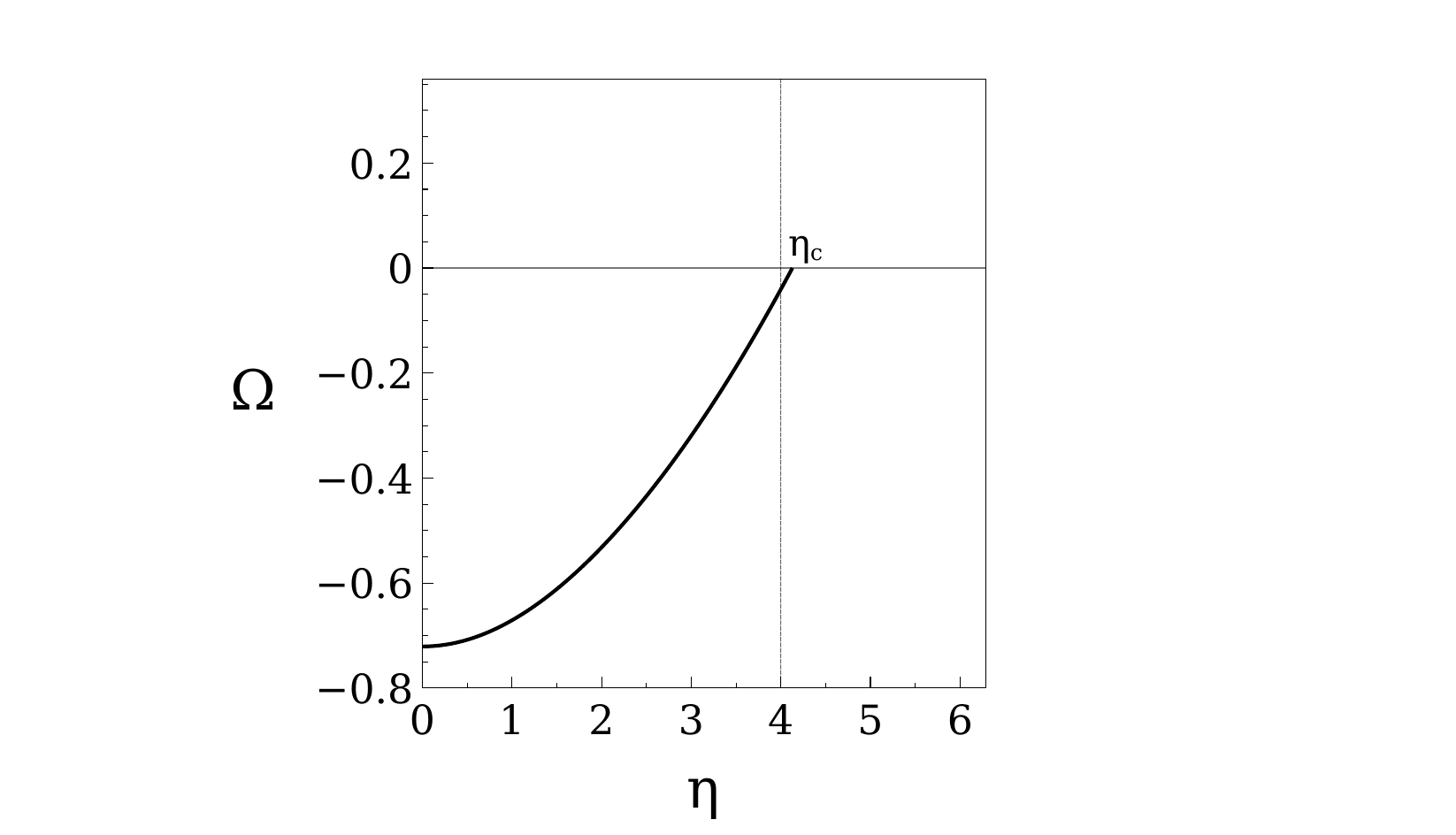}
\end{center}
\vspace{-0.5cm}
\caption{Frequency $\Omega=$ Arg\! $Q_1$ versus alignment noise $\eta$ along the critical line $p(\eta)$ that separates Regions I (disordered phase) and III (rotating wave phase) in the lower part of Fig.~\ref{fig2}(a). Increasing $\eta$ corresponds to decreasing $p$ along this critical line. Parameter values are $\xi_0=3\pi/4$ and $M=7$. \label{fig3}}
\end{figure}

\subsection{Phase diagrams}
\subsubsection{Noises $p$ vs $\eta$} Fig.~\ref{fig2} depicts the stability regions of the disordered state in the parameter space $(p,\eta)$ at zero wave number. $Q_1$ given by Eq.~\eqref{eq20} is real if $g(\xi)$ is even. For $\xi_0=\pi$ in Fig.~\ref{fig2}(b), we shall see in Section \ref{sec:6} that there are supercritical pitchfork bifurcations at the critical line I-II ($Q_1=1$) and supercritical period doubling bifurcations at the critical line I-IV ($Q_1=-1$). If $p>0$ and $0<\xi_0<\pi$ in Eq.~\eqref{eq3}, {\em the noise density is no longer even}, $Q_1$ is complex, and the order-disorder phase transition occurs with $|Q_1|=1$. The eigenfunction is a rotating wave, $\tilde{f}(\theta,t)=e^{i(\theta+\Omega t)}$, $\Omega=\text{Arg }Q_1$. Other modes have $|Q_j|<1$ and decay as $t\to\infty$. There are supercritical Hopf bifurcations at the critical lines separating the Regions I and III in Fig.{\ref{fig2}(a); cf. Section~\ref{sec:5}. 

Note that increasing the angle $\xi_0$ makes Region I larger. Fig.~\ref{fig2}(a) shows the critical lines separating Region I (stable incoherent motion) from rotating wave phases (RWPs) in Region III: as $\xi_0$ approaches the value $\pi$, the lower critical line moves down toward the line separating Regions I and II in Fig.~\ref{fig2}(b) and $\Omega=$ Arg\! $Q_1$ approaches zero. Meanwhile, the upper critical line in Fig.~\ref{fig2}(a) moves up towards the line separating Regions I and IV in Fig.~\ref{fig2}(b) and $\Omega$ approaches $-1$. As the phase diagrams and numerical simulations of the VM are similar for $\pi/2<\xi_0<\pi$, we have selected $\xi_0=3\pi/4$ to present our results for RWPs. By an abuse of language, we will call this the case of {\em almost contrarian compulsions} (even though $\xi_0=3\pi/4$ is not close to contrarian compulsions with $\xi_0=\pi$). Fig.~\ref{fig3} depicts the frequency $\Omega$ as a function of alignment noise on the line $\eta(p)$ that separates Regions I and III in the lower part of Fig.~\ref{fig2}(a) (for $\xi_0=3\pi/4$). Note that $\Omega=0$ at the critical value $\eta_c$ of the standard VM corresponding to $p=0$.

\begin{figure}[h]
\begin{center}
\includegraphics[width=8.5cm]{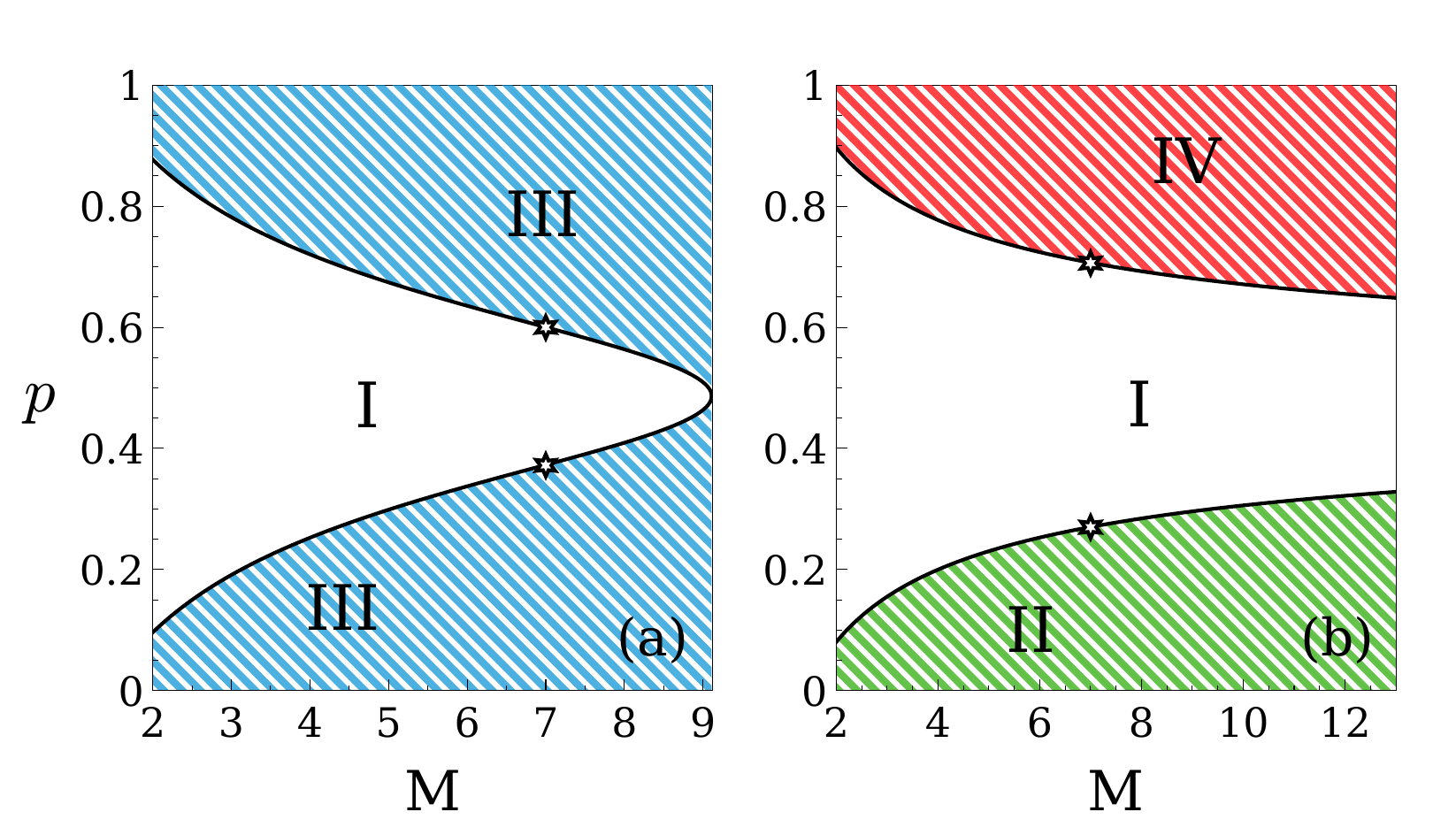}
\end{center}
\caption{Phase diagram of $p$ versus $M$ obtained from kinetic theory. Stable phases at each region are: I (incoherent), II (stationary), III (rotating wave), IV (period 2). Parameter values are $\eta=\pi/3$, and: (a) $\xi_0=3\pi/4$ (solid line) and (b) $\xi_0=\pi$. The value $M=7$ used in Fig.~\ref{fig3} is marked by a star. \label{fig4}}
\end{figure}

\subsubsection{Average number of neighbors $M$ vs $p$}
The other parameter appearing in Eq.~\eqref{eq20} is the average number of neighboring particles, $M=\rho_0\pi R_0^2$. This parameter changes with the radius of the influence region or the average number density. Fig.~\ref{fig4} shows the phase diagram of $p$ versus $M$ for a typical value of alignment noise, $\eta=\pi/3$. Increasing the average number of particles inside the circle of influence favors swarming phases, and therefore Regions II, III and IV (polarized phases) grow at the expense of Region I (zero polarization). For $\xi_0=3\pi/4$, Fig.~\ref{fig4}(a) shows that Region I disappears for $M$ larger than a critical value of about 9. When the average number of neighbors is larger than this critical value, $M_*(\xi_0)$, particles are always polarized. $M_*(\xi_0)$ increases with the angle $\xi_0$ for a fixed value of the noise strength $\eta$. Numerical simulations suggest that $M_*(\pi)=+\infty$.

\section{Bifurcation theory} \label{sec:4}
The case of critical eigenvalue $Q_1=1$ (pitchfork bifurcation) has been discussed in Ref.~\cite{BT18}. Here we shall use the same method to analyze the bifurcations at zero wave number when the critical eigenvalues are  $Q_1=-1$ (period doubling bifurcation) and complex $Q_1=e^{i\Omega}$ (Hopf bifurcation). We shall use the alignment noise $\eta$ as a bifurcation parameter and comment on the change in our results had the average particle density been used instead. The solution of the linearized equation 
\begin{eqnarray}
\mathcal{L}f^{(1)}\!\!\equiv\! f^{(1)}(\theta,t+1,\textbf{X},T)\!-\! C^{(1)}[ f^{(1)}](\theta,t,\textbf{X},T)\!=\!0,\label{eq24}
\end{eqnarray}
is  
\begin{eqnarray}
 f^{(1)}(\theta,t,\textbf{X},T,\epsilon)\!=\!\frac{r(\textbf{X},T;\epsilon)}{2\pi}\!+\!A(\textbf{X},T;\epsilon) e^{i(\theta+\Omega t)} \!+\!\mbox{cc},\label{eq25}\\
 \Omega=\text{Arg }Q_1,\quad \textbf{X}=\epsilon\textbf{x},\quad T=\epsilon t.\quad \label{eq26}
\end{eqnarray}
Here cc means the complex conjugate of the preceding term. We do not need to include more terms in Eq.~\eqref{eq25} because the other modes decay rapidly in the fast time scale $t$. The argument of the eigenvalue $Q_1$ is $\Omega=0$, (pitchfork bifurcation), $\Omega=\pi$ (period doubling bifurcation) and $\Omega\in(-\frac{\pi}{2},0)$ (Hopf bifurcation). The two latter cases can be treated together ($\Omega\neq 0$).

Bifurcation theory is quite different in the cases of uniform and space dependent density. The case of uniform density can be treated by using multiple scales for all bifurcation types: pitchfork (I-II), Hopf (transition I-III) and period doubling (I-IV); see Appendix \ref{ap:2}. Here we consider space dependent density. We anticipate crossover scalings and therefore we shall use the Chapman-Enskog method \cite{BT10}. In all three bifurcation cases, the hierarchy of equations resulting from the Chapman-Enskog ansatz \cite{BT10,ace05,BT18}, 
\begin{eqnarray}
f(\theta,\textbf{x},t;\epsilon)=f_0+ \epsilon f^{(1)}+\sum_{j=2}^\infty\epsilon^j f^{(j)}(\theta,t;r,A,\overline{A}), \label{eq27}\\
\frac{\partial r}{\partial T}=\mathcal{R}^{(0)}(r,A,\overline{A})+\epsilon\mathcal{R}^{(1)}(r,A,\overline{A})+O(\epsilon^2),\label{eq28}\\
\frac{\partial A}{\partial T}=\mathcal{A}^{(0)}(r,A,\overline{A})+\epsilon\mathcal{A}^{(1)}(r,A,\overline{A})+O(\epsilon^2),\label{eq29}
\end{eqnarray}
with $\eta=\eta_c+\epsilon^2\eta_2$ ($\overline{A}$ is the complex conjugate of $A$), is
\begin{eqnarray}
\mathcal{L}f^{(2)}=C^{(2)}[f^{(1)},f^{(1)}]-\textbf{v}\!\cdot\!\nabla_X f^{(1)}|_{t+1}- \frac{\mathcal{R}^{(0)}}{2\pi}\nonumber\\
-\mathcal{A}^{(0)} e^{i(\theta+\Omega t+\Omega)}+ \mbox{cc},\label{eq30}
\end{eqnarray}
\begin{eqnarray}
&&\mathcal{L}f^{(3)}=C^{(3)}[f^{(1)},f^{(1)},f^{(1)}]+2 C^{(2)}[f^{(1)},f^{(2)}]  \nonumber\\
&&\, -\!\left(\frac{\partial}{\partial T}+\textbf{v}\!\cdot\!\nabla_X\!\right)\! f^{(2)}|_{t+1}-\frac{\mathcal{R}^{(1)}}{2\pi}-\mathcal{A}^{(1)} e^{i(\theta+\Omega t+\Omega)}
\nonumber\\
&&\,+ \mbox{cc}\!-\frac{1}{2}\!\left(\frac{\partial}{\partial T}\!+\!\textbf{v}\!\cdot\!\nabla_X\!\right)^2\!\!f^{(1)}|_{t+1}\! +\eta_2\frac{\partial}{\partial\eta}C^{(1)}[f^{(1)}],\quad\label{eq31}
\end{eqnarray}
etc. In these equations, $C^{(1)}[\varphi]$ is given by Eq.~\eqref{eq16} with $\mathbf{K}=0$, and we have 
\begin{widetext}
\begin{eqnarray}
&&C^{(2)}[\varphi,\varphi]=\frac{\pi R_0^2}{2}\!\left[\sum_{n=2}^\infty\frac{ne^{-M}}{(n-2)!}\!\left(\frac{M}{2\pi}\right)^{n-2}\!\int_{-\pi}^\pi d\xi\, g(\xi)\int_{[-\pi,\pi]^n}\hat{\delta}(\theta-\xi-\Phi_1)\varphi(\theta_1)\varphi(\theta_2)\prod_{l=1}^nd\theta_l\right.\nonumber\\
&&\left.-2\left(\int_{-\pi}^\pi \varphi(\theta_1)d\theta_1\right)\!\sum_{n=1}^\infty\frac{ne^{-M}}{(n-1)!}\!\!\left(\frac{M}{2\pi}\right)^{n-1}\!\int_{-\pi}^\pi d\xi\, g(\xi)\!\int_{[-\pi,\pi]^n}\!\hat{\delta}(\theta-\xi-\Phi_1)\varphi(\theta_1)\!\prod_{l=1}^nd\theta_l\! 
+\frac{M}{2\pi}\left(\int_{-\pi}^\pi \varphi(\theta_1)d\theta_1\right)^2\right]\!,\quad\label{eq32}
\end{eqnarray}
\begin{eqnarray}
&&C^{(3)}[\varphi,\varphi,\varphi]=\frac{\pi^2R_0^4}{6}\!\left[\sum_{n=3}^\infty\frac{ne^{-M}}{(n-3)!}\!\left(\frac{M}{2\pi}\right)^{n-3}\!\int_{-\pi}^\pi d\xi\, g(\xi)\int_{[-\pi,\pi]^n}\!\hat{\delta}(\theta-\xi-\Phi_1)\varphi(\theta_1)\varphi(\theta_2)\varphi(\theta_3)\!\prod_{l=1}^nd\theta_l\! \right.\nonumber\\
&&-3\!\left(\int_{-\pi}^\pi \varphi(\theta_3)d\theta_3\right)\!\sum_{n=2}^\infty\frac{ne^{-M}}{(n-2)!}\!\left(\frac{M}{2\pi}\right)^{n-2}\!\int_{-\pi}^\pi d\xi\, g(\xi)\int_{[-\pi,\pi]^n}\!\hat{\delta}(\theta-\xi-\Phi_1)\varphi(\theta_1)\varphi(\theta_2)\!\prod_{l=1}^nd\theta_l\! \nonumber\\
&&+3\!\left(\int_{-\pi}^\pi \varphi(\theta_2)d\theta_2\right)^2\!\sum_{n=1}^\infty\frac{ne^{-M}}{(n-1)!}\!\left(\frac{M}{2\pi}\right)^{n-1}\!\int_{-\pi}^\pi d\xi\, g(\xi)\int_{[-\pi,\pi]^n}\!\hat{\delta}(\theta-\xi-\Phi_1)\,\varphi(\theta_1)\!\prod_{l=1}^nd\theta_l\! 
-\left.\frac{M}{2\pi}\left(\int_{-\pi}^\pi \varphi(\theta_1)\, d\theta_1\right)^3\right]\!,\quad\label{eq33}
\end{eqnarray}
\end{widetext}
and so on. Note that $C_E[f_0+\epsilon\tilde{\rho}]=f_0+\epsilon\tilde{\rho}$ and $C^{(1)}[\tilde{\rho}]=\tilde{\rho}$ for constant $\tilde{\rho}$ imply $C^{(2)}[1,1]=C^{(3)}[1,1,1]=0$, which can be checked from Eqs.~\eqref{eq32}-\eqref{eq33}. The solvability conditions for non-homogeneous equations of the hierarchy is that their right hand sides be orthogonal to the solutions of the homogeneous equation $\mathcal{L}\varphi=0$, namely 1 and $e^{i(\theta+\Omega t)}$, using the scalar product
\begin{equation}
\langle f(\theta,t),g(\theta,t)\rangle=\lim_{S\to\infty}\frac{1}{S}\sum_{t=0}^S \int_{-\pi}^\pi \overline{f(\theta,t)} g(\theta,t)\, d\theta. \label{eq34}
\end{equation}
These solvability conditions applied to Eqs.~\eqref{eq30} and \eqref{eq31} yield the terms appearing in Eqs.~\eqref{eq28} and \eqref{eq29}, which are the amplitude equations. In the next two sections, we consider separately the three bifurcation types.

\section{Hopf and period doubling bifurcations}\label{sec:5}
For $0<p\leq 1$ and $0<\xi_0<\pi$, we have $Q_1=e^{i\Omega}$, with $-\pi/2\leq\Omega <0$ in Eq.~\eqref{eq26}, for a critical noise value $\eta_c$ located on either the upper or the lower critical lines separating Regions III from Region I in Fig.~\ref{fig2}(a). Similarly, for $\xi_0=\pi$, $p>p_0\approx 2/\sqrt{\pi M}$ [$p_0$ corresponds to $\eta_c=2\pi$, $Q_1=-1$ in Eq.~\eqref{eq20} for $\xi_0=\pi$], we have $\Omega=\pi$ at the critical noise $\eta_c$ separating Regions I and IV in Fig.~\ref{fig2}(b). Setting a nonzero $\Omega$, we can treat these two cases simultaneously. The solvability conditions for Eq.~\eqref{eq30} are that its right hand side should be orthogonal to the solutions of the homogeneous equation $\mathcal{L}\varphi(\theta)=0$, which are 1 and $e^{i(\theta+\Omega t)}$. Using the scalar product \eqref{eq34}, these conditions yield
\begin{eqnarray}
\mathcal{R}^{(0)}=0,\quad \mathcal{A}^{(0)}=\frac{C^{(2)}[1,e^{i\theta}]_1}{\pi}rAe^{-i\Omega}\sim \frac{rA}{2\rho_0}.\label{eq35}
\end{eqnarray}
Eq.~\eqref{eq30} has the solution
\begin{eqnarray}
f^{(2)}(\theta,t,\textbf{X},T)\!=\!\! \left[\frac{A^2 e^{i2(\theta+\Omega t)}C^{(2)}[e^{i\theta},e^{i\theta}]_2}{e^{i2\Omega}-C^{(1)}[e^{i2\theta}]_2} \right. \quad\quad\quad\quad\nonumber\\
-\frac{e^{i(2\theta+\Omega t)}\!\left(\frac{\partial}{\partial X} - i\frac{\partial}{\partial Y}\right)\!A}{2(e^{i\Omega}-C^{(1)}[e^{i2\theta}]_2)} \!-\! \frac{e^{i\Omega(t+1)}\!\left(\frac{\partial}{\partial X} + i\frac{\partial}{\partial Y}\right)\!A}{2(e^{i\Omega}-1)}\nonumber\\
\!+\!\left.\frac{e^{i\theta}\!\left(\frac{\partial}{\partial X} - i\frac{\partial}{\partial Y}\right)\! r}{2(e^{i\Omega}-1)}\right]\!+\!\mbox{cc}. \quad\label{eq36}
\end{eqnarray}
Inserting Eqs.~\eqref{eq25} and \eqref{eq36} in Eq.~\eqref{eq31} and using the solvability conditions, we find $\mathcal{R}^{(1)}=0$ and $\mathcal{A}^{(1)}$. Then, up to terms of order $\epsilon^2$, the amplitude equations are 
\begin{eqnarray}
&&\frac{\partial r}{\partial T}=0,\label{eq37}\\
&&\frac{\partial A}{\partial T}=\!\left[\frac{r}{2\rho_0}+\epsilon\!\left(\eta_2Q_\eta-\frac{r^2e^{i2\Omega}}{8\rho_0^2}-\mu|A|^2\right)\!\right]\!A\quad\nonumber\\
&&\quad\quad+\,\epsilon\delta\nabla_X^2A,\label{eq38}
\end{eqnarray}
in the limit as $M\gg 1$. In the same limit we also have
\begin{eqnarray}
Q_\eta\!&=&\!\frac{\partial}{\partial\eta}\!\left(\ln C^{(1)}[e^{i\theta}]_1\right)\big|_{\eta_c}\!\nonumber\\
\!&\sim&\!\frac{\sqrt{\pi M}}{\eta_c^2}(1-p)e^{-i\Omega}\!\left(\frac{\eta_c}{2}\cos\frac{\eta_c}{2}-\sin\frac{\eta_c}{2}\right)\!,\label{eq39}
\end{eqnarray}
and we prove in Appendix \ref{ap:3} that $\delta$ and $\mu$ both have positive real parts. According to Eq.~\eqref{eq37}, $r$ is independent of $T$. Conservation of the number of particles implies
\begin{equation}
\int r(\mathbf{X})\, d\mathbf{X}=0. \label{eq40}
\end{equation}

\subsection{Complex Ginzburg-Landau equation for rescaled $r$}
For $\epsilon=0$, the solution of Eq.~\eqref{eq38} yields an $|A|$ that increases in time for $r(\mathbf{X})>0$ and it decreases for $r(\mathbf{X})<0$. This indicates that a dominant balance occurs only if we assume $r(\mathbf{X})=O(\epsilon)$. Then Eq.~\eqref{eq38} becomes a modified complex Ginzburg-Landau equation (CGLE) with a diffusive scaling for the time:
\begin{eqnarray}
\frac{\partial A}{\partial(\epsilon^2t)}=\delta\,\nabla^2_X A + \left[\eta_2 Q_\eta+ \frac{r(\mathbf{X})}{2\rho_0} - \mu |A|^2\right] A. \label{eq41}
\end{eqnarray}
If the density is kept uniform, $r=0$, Eq.~\eqref{eq41} is the usual CGLE. In this case, it has the rotating wave solution 
\begin{eqnarray}
&&A(\epsilon^{2}t)=\sqrt{\frac{\eta_2\mbox{Re}Q_\eta}{\mbox{Re}\mu}}\, e^{i\eta_2\varphi\epsilon^{2}(t-t_0)},\label{eq42}\\ 
&&\varphi=\mbox{Im}Q_\eta-\mbox{Im}\mu\,\frac{\mbox{Re}Q_\eta}{\mbox{Re}\mu}.   \label{eq43}
\end{eqnarray}
As proven in Appendix \ref{ap:3}, Re$\mu>0$, and therefore the phases issuing forth from Hopf and period doubling bifurcations are both supercritical: they exist only for $\eta<\eta_c$ (where the uniform distribution is unstable) and are linearly stable against space-independent disturbances. The polarization corresponding to the bifurcating solutions given by Eqs.~\eqref{eq42}-\eqref{eq43} is the modulus of the complex parameter: 
\begin{eqnarray}
Z&=&\frac{1}{N}\sum_{j=1}^N e^{i\theta_j}=\frac{1}{N}\int e^{i\theta}f(\theta,\mathbf{x},t)\, d\theta d\mathbf{x}\nonumber\\
&\sim&\frac{2\pi}{\rho_0} \sqrt{\frac{(\eta-\eta_c)\,\mbox{Re}Q_\eta}{\mbox{Re}\mu}}\, e^{i\theta_0-i[\Omega+(\eta-\eta_c)\varphi]t}.  \label{eq44}
\end{eqnarray}
Near $\eta_c$, we have $Q_1\sim e^{i\Omega}[1+ Q_\eta(\eta-\eta_c)]$. Then we can replace $e^{-i\Omega}Q_1-1$ instead of $Q_\eta(\eta-\eta_c)$ in Eq.~\eqref{eq44}, thereby obtaining a formula that holds for larger values of $|\eta-\eta_c|$:
\begin{eqnarray}
&&Z\sim\frac{2\pi}{\rho_0} \sqrt{\frac{\mbox{Re}(e^{-i\Omega}Q_1)-1}{\mbox{Re}\mu}}\, e^{i\theta_0-i(\Omega+\psi) t},  \label{eq45}\\
&&\psi= \mbox{Im}(e^{-i\Omega}Q_1)- \frac{\mbox{Re}(e^{-i\Omega}Q_1)-1}{\mbox{Re}\mu}\,\mbox{Im}\mu. \label{eq46}
\end{eqnarray}

The bifurcating solutions given by Eqs.~\eqref{eq42}-\eqref{eq43} are uniform in space. However, Eq.~\eqref{eq38} is a nonlinear reaction-diffusion equation with a diffusion coefficient whose real part is positive according to Eq.~\eqref{a3_6}. For a nonuniform particle density, Eq.~\eqref{eq40} (conservation of the number of particles) strongly suggests the formation of ordered clusters. Let us imagine that $r(\mathbf{X})=s^2\rho_0$sign$(X-L/2)$, $s^2>\eta_2$Re$Q_\eta$. Then $W=0$ for $X>L/2$, whereas for $X<L/2$ we have $W>0$, given approximately by Eq.~\eqref{eq42} with $\eta_2$Re$Q_\eta+s^2$ instead of $\eta_2$Re$Q_\eta$. The phase of the complex order parameter satisfies the integrated form of the Burgers equation \cite{kur76}
\begin{eqnarray}
&&\frac{\partial\Upsilon}{\partial\sigma}=\kappa\nabla_X^2\Upsilon-\frac{1}{2}|\nabla_X\Upsilon|^2,\label{eq47}\\
&& \Upsilon=2\left(\frac{\mbox{Im}\delta}{\mbox{Re}\delta}-\frac{\mbox{Im}\mu}{\mbox{Re}\mu}\right)\mbox{Arg}(A),\label{eq48}\\
&&\sigma=\epsilon^2\mbox{Re}\delta\, t,\quad\kappa=1+\frac{\mbox{Im}\delta\,\mbox{Im}\mu}{\mbox{Re}\delta\,\mbox{Re}\mu}. \label{eq49}
\end{eqnarray}
Near the bifurcation line I-III in Fig.~\ref{fig2}(a), $\kappa>0$. For the geometry we are considering, $\Upsilon$ depends only on the coordinate $X$ and on $\sigma$. Then $k=\partial\Upsilon/\partial X$ satisfies the Burgers equation proper. Assuming that $L\gg 1$, $k\sim k_2>0$ at $X=0$ and $k=0$ at $X=L/2$, $k$ is the shock wave solution given by Eq.~(4.23) of Ref.~\cite{whi74}:
\begin{eqnarray}
\frac{\partial\Upsilon}{\partial X}=\frac{k_2}{1+\exp\left[\frac{k_2}{2\kappa}(X-\frac{k_2}{2}\sigma-X_0)\right]}. \label{eq50}
\end{eqnarray}
This solution represents a planar wave front moving to the right with velocity $k_2/2$. The front  encroaches an unpolarized region with zero wave number and leaves a polarized region with wave number $k_2$ on its wake. The region behind the wave front is a cluster rotating with angular velocity $\Omega$ and local wave number proportional to $k_2$. This simple example illustrates how a non-constant density may produce inhomogeneous ordered clusters of the RWP.

Eq.~\eqref{eq42} and related plane wave solutions of the CGLE for $r(\mathbf{X})=0$ become unstable provided $\kappa<0$ in Eq.~\eqref{eq47}. Equivalently, $b_3<b_1$,  $b_1\!=$ Im$\delta/$Re$\delta$ and $b_3=-$Re$\mu/$Im$\mu$ (Newell's criterion). Close to the line $b_1=b_3$, one can derive the Kuramoto-Sivashinsky equation, \cite{cha96,ara02,man04}. Phase turbulence consisting of disordered cellular structures is then possible. Solutions of the CGLE with periodic boundary conditions also include spiral waves and other defects (having $W=0$ at one point in their cores), as well as phases of defect turbulence in which defects are created and annihilated in pairs \cite{cha96}. For the period 2 solution, Eq.~\eqref{eq25} with $\Omega=\pi$, $\delta$ and $\mu$ are real. Then Eq.~\eqref{eq41} has vortex solutions with nonzero rotation number and a vortex gas evolves as indicated in Refs.~\cite{neu90,cro93,cha96,ara02,man04}. 

\subsection{Complex Ginzburg-Landau equation for random density disturbances}
Let us assume now that $\rho(\mathbf{X})-\rho_0=\epsilon r(\mathbf{X})$ is a zero-mean random Gaussian process with standard deviation $\sigma_r=\epsilon\tilde{\sigma}_r$. Then the average value  $\langle r A\rangle\approx \langle r\rangle\langle A\rangle=0$, and the mean amplitude, $\tilde{A}=\langle A\rangle$, satisfies the approximate equation:
\begin{eqnarray}
\frac{\partial\tilde{A}}{\partial(\epsilon^2t)}=\delta\,\nabla^2_X\tilde{A} + \left[\eta_2 Q_\eta- \frac{\tilde{\sigma}_r^2}{8\rho_0^2} -\mu |\tilde{A}|^2\right] \tilde{A}. \label{eq51}
\end{eqnarray}
Here we have made $\langle r^2A\rangle\approx \tilde{\sigma}_r^2\tilde{A}$, $\langle|A|^2A\rangle\approx |\tilde{A}|^2\tilde{A}$. Now the uniform solution is Eq.~\eqref{eq42} in which $\eta_2\mbox{Re}Q_\eta$ is replaced by $[\eta_2\mbox{Re}Q_\eta-\tilde{\sigma}_r^2/(8\rho_0^2)]$. Then Re$(e^{-i\Omega}Q_1)-1$ in Eq.~\eqref{eq45} for the order parameter is replaced by [Re$(e^{-i\Omega}Q_1)-1- \sigma_r^2/(8\rho_0^2)]$, with the result 
\begin{equation}
W=|Z|\sim\frac{2\pi}{\rho_0}\sqrt{\frac{\mbox{Re}(e^{-i\Omega}Q_1)-1-\frac{\sigma_r^2}{8 \rho_0^2}}{\mbox{Re}\mu}}. \label{eq52}
\end{equation}
Equating to zero this last quantity, we find the critical value of the noise, $\eta_c$, which gets shifted to a smaller value. How large is $\sigma_r^2=\langle r^2\rangle$? We know that the particles are randomly placed in the box at the initial time. The fluctuation of the density is 
\begin{eqnarray}
\langle (\rho-\rho_0)^2\rangle=\frac{\rho_0^2}{N} \frac{k_BT}{-v^2\frac{\partial P}{\partial v}}=\frac{\rho_0^2}{N}, \label{eq53}
\end{eqnarray}
where $k_B$, $T$, $P$, and $v=1/\rho_0$ are the Boltzmann constant, temperature, pressure and specific volume, respectively [cf. Eq.~(7.43) in Ref.~\cite{hua87}]. The particles can be thought of as belonging to an ideal gas at the initial time, therefore $P=\rho_0k_BT$ and $-v^2\partial P/\partial v=\partial P/\partial\rho_0=k_BT$. Then we have $\langle (\rho-\rho_0)^2\rangle=\rho_0^2/N$, as written in Eq.~\eqref{eq53}, and Eq.~\eqref{eq52} becomes
\begin{equation}
W=|Z|\sim\frac{2\pi}{\rho_0}\sqrt{\frac{\mbox{Re}(e^{-i\Omega}Q_1)-1-\frac{1}{8N}}{\mbox{Re}\mu}}.\label{eq54}
\end{equation}
The shift in the bifurcation point indicated by Eq.~\eqref{eq54} vanishes in the limit as $N\to\infty$. 

\subsection{Average particle density as bifurcation parameter}
What happens if we select the average particle density $\rho_0$ as bifurcation parameter instead of the alignment noise $\eta$? 

Firstly, we have to replace $\rho_0=\rho_{0c}+\epsilon^2\rho_2$ instead of $\eta=\eta_c+\epsilon^2\eta_2$ and $f^{(2)}+\rho_2$ instead of $f^{(2)}$ in Eq.~\eqref{eq27}. Then a term $\rho_2\pi R_0^2\partial C^{(1)}[f^{(1)}]/\partial M$ replaces the last term in the right hand side of Eq.~\eqref{eq31}, to which we have to add a term $2\rho_2 C^{(2)}[1,f^{(1)}]$. For all bifurcation types, $\rho_2Q_\rho$ replaces $\eta_2Q_\eta$ in the amplitude equations. $Q_\rho$ is defined as the derivative with respect to $\rho_0$ instead of the derivative with respect to $\eta$ in the first line of Eq.~\eqref{eq39}. Secondly, for Hopf or period doubling bifurcations, we replace  $r(\mathbf{X})+2\pi\rho_2$ instead of $r(\mathbf{X})$ in Eqs.~\eqref{eq40} and \eqref{eq41}. The remaining considerations are the same provided we make these replacements.

One obvious change is that Re$Q_\rho>0$ (cf. Fig~\ref{fig4}), whereas Re$Q_\eta<0$ on the lower critical lines of Fig~\ref{fig2}, and Re$Q_\eta>0$ on the upper critical lines of the same figure. Thus, the disordered phase is stable for $\rho_0<\rho_{0c}$ and unstable for  $\rho_0>\rho_{0c}$, whereas the situation is the opposite for the polarized random wave and stationary phases on the lower sectors of Fig~\ref{fig2} if we use the alignment noise as a bifurcation parameter. 

\section{Pitchfork bifurcation}\label{sec:6}
For $p=0$ (standard VM) or $p>0$ and $\xi_0=\pi$, corresponding to the lower sector of the phase diagram in Fig. \ref{fig2}(b), the critical condition is $Q_1=1$, and therefore $\Omega=0$ in Eq.~\eqref{eq26}. The corresponding bifurcation has been analyzed in Ref.~\cite{BT18} for the case $p=0$. The same procedure based on the solvability conditions for Eqs.~\eqref{eq30} and \eqref{eq31} produces nonzero $R^{(0)}$ and $R^{(1)}$ and the density disturbance is no longer time independent. The amplitude equations are equivalent to the following system for the density disturbance and a current density $\mathbf{w}=2\pi(\mbox{Re}A,-\mbox{Im}A)$:
\begin{eqnarray}
&&\frac{\partial r}{\partial T}+\nabla_X\!\cdot\!\left[\!\left(1+\frac{\epsilon\gamma_3r}{\pi}\right)\!\mathbf{w}\right]\!=0,\label{eq55}\\
\frac{\partial \mathbf{w}}{\partial T}\!&\!=\!&\!-\frac{1}{2}\nabla_X\!\left[\!\left(1-\frac{\epsilon\gamma_3}{4\pi}r\right)\!r+\frac{\epsilon(2\gamma_1-\gamma_2-\gamma_3)}{2\pi}|\mathbf{w}|^2\!\right]\! \nonumber\\
&\!+\!& \!\epsilon\frac{2\gamma_1-\gamma_2+\gamma_3}{2\pi}\mathbf{w}(\nabla_X\!\cdot\mathbf{w})  \nonumber\\
&\!+\!&\!\epsilon\frac{2\gamma_1+\gamma_2-\gamma_3}{2\pi}(\mathbf{w}\cdot\nabla_X)\mathbf{w}+\epsilon\delta\nabla^2_X \mathbf{w} \nonumber\\
&\!+\!&\!\!\left[\frac{2\gamma_3 r}{\pi}+\epsilon\!\left(\eta_2Q_\eta-\frac{r^2}{6\rho_0}\!-\!\frac{\mu |\mathbf{w}|^2}{4\pi^2}\right)\!\right]\! \mathbf{w}. \label{eq56}
\end{eqnarray}
The coefficients appearing in these equations are all real valued and listed in Appendix \ref{ap:3}. For $r=0$ and $\gamma_3= 0$, these equations are exactly equivalent to the amplitude equation (132) of Ref.~\cite{ihl16}. As in the case of Eq.~\eqref{eq38}, $\delta$ and $\mu$ are both positive. For $\epsilon=0$, Eq.~\eqref{eq55} is the continuity equation for a density variable $r$ and a current density $\mathbf{w}$, which explains the name of the latter variable. The overall density of particles is $N/L^2$, which implies the following constraint for $r(\mathbf{X},T)$:
\begin{equation}
\int r(\mathbf{X},T)\, d\mathbf{X}=0.\label{eq57}
\end{equation}

\subsection{Space independent $A$ and $r=0$: diffusive scaling}
For space independent $A$, $r=0$, and Eq.~\eqref{eq56} is the typical pitchfork amplitude equation with diffusive scaling. It has a stationary solution with modulus
\begin{equation}
|\mathbf{w}_0|=2\pi\sqrt{\frac{\eta_2Q_\eta}{\mu}},\label{eq58}
\end{equation}
which is stable, and it exists for $\eta<\eta_c$, $\mu>0$, and $Q_\eta<0$. For $\eta<\eta_c$,  the uniform distribution $f_0$ is unstable as $Q_1>1$ there. Thus the transition from incoherence to order is a supercritical pitchfork bifurcation. The polarization corresponding to this solution is 
\begin{eqnarray}
W_0\sim\frac{2\pi}{\rho_0} \sqrt{\frac{(\eta-\eta_c)Q_\eta}{\mu}}.  \label{eq59}
\end{eqnarray}
Near $\eta_c$, we have $Q_1\sim 1+ Q_\eta(\eta-\eta_c)$. Then we can replace $Q_1-1$ instead of $Q_\eta(\eta-\eta_c)$ in Eq.~\eqref{eq59}, thereby obtaining a formula that holds for larger values of $|\eta-\eta_c|$:
\begin{eqnarray}
W_0\sim\frac{2\pi}{\rho_0} \sqrt{\frac{Q_1-1}{\mu}}. \label{eq60}
\end{eqnarray}

\subsection{Convective scaling and resonance}
This case has been analyzed in Ref.~\cite{BT18}. In this paper, we describe the main results and line of argumentation found in Ref.~\cite{BT18} for the sake of completeness. Close to the bifurcation point, $\epsilon\ll 1$, the diffusive and convective scalings are well separated. Then the solution of the amplitude equations produce a polarization close to that in Eqs.~\eqref{eq59}-\eqref{eq60} but there are persistent oscillations about it in the convective time scale. The leading order Eqs.~\eqref{eq55}-\eqref{eq56} for $\epsilon=0$ and linearized about Eqs~\eqref{eq58} are 
\begin{eqnarray}
&&\frac{\partial\tilde{r}}{\partial T}+\nabla_X\!\cdot\!\mathbf{\tilde{w}}=0,\label{eq61}\\
&&\frac{\partial \mathbf{\tilde{w}}}{\partial T}=-\frac{1}{2}\nabla_X\tilde{r}+\frac{\gamma_3}{\pi}\mathbf{w}_0\tilde{r}. \label{eq62}
\end{eqnarray}
By differentiating Eq.~\eqref{eq61} and eliminating $\mathbf{\tilde{w}}$ by means of Eq.~\eqref{eq62}, we find the wave equation:
\begin{equation}
\frac{\partial^2\tilde{r}}{\partial T^2}=\frac{1}{2}\nabla_X^2\tilde{r} -\frac{\gamma_3}{\pi}\, \mathbf{w}_0\cdot\nabla_X\tilde{r}. \label{eq63}
\end{equation}
The change of variable
\begin{equation}
\tilde{r} = e^{\gamma_3\mathbf{w}_0\cdot\mathbf{X}/\pi} R, \label{eq64}
\end{equation}
eliminates the gradient term in Eq.~\eqref{eq63}, thereby producing the Klein-Gordon equation:
\begin{equation}
\frac{\partial^2R}{\partial T^2}=\frac{1}{2}\nabla_X^2R -\frac{\gamma_3^2|\mathbf{w}_0|^2}{2\pi^2}\, R.\label{eq65}
\end{equation}
For periodic boundary conditions, $R(\mathbf{X},T)$ can be expanded in a Fourier series
\begin{eqnarray}
&&R(\mathbf{X},T)=\sum_{n=1}^\infty\sum_{m=1}^\infty R_{n,m} e^{i\mathbf{k}_{n,m}\cdot\mathbf{X}},\label{eq66}\\
&& \mathbf{k}_{n,m}=\frac{2\pi}{\epsilon L}(n,m). \label{eq67}
\end{eqnarray}
Inserting this equation into Eq.~\eqref{eq65}, we obtain the equation of a linear oscillator:
\begin{eqnarray}
&&\frac{d^2R_{n,m}}{dT^2}+\omega_{n,m}^2R_{n,m}=0, \label{eq68}\\
&&\omega_{n,m}=\sqrt{\frac{1}{2}|\mathbf{k}_{n,m}|^2 +\frac{\gamma_3^2|\mathbf{w}_0|^2}{2\pi^2}}. \label{eq69}
\end{eqnarray}
In the original time scale, these frequencies are
\begin{equation}
\epsilon\,\omega_{n,m}=\sqrt{\frac{2\pi^2}{L^2}(n^2+m^2) +\frac{W_0^2}{8}}, \label{eq70}
\end{equation}
where $W_0$ is the polarization of Eq.~\eqref{eq60}. Formulas for $\tilde{r}(\mathbf{X},T)$ and $\mathbf{\tilde{w}}(\mathbf{X},T)$ can be found in Ref.~\cite{BT18}. The contributions of Fourier modes having nonzero frequency to the polarization can be made to resonate with an external forcing added to the alignment rule of Eq.~{\eqref{eq1} (cf. Ref.~\cite{BT18}). 

The oscillating density disturbance $\tilde{r}(\mathbf{X},T)$ produces a nonzero value of the average of $r^2=\tilde{r}^2$ over space and time. The polarization becomes \cite{BT18}:
\begin{eqnarray}
W\!&\sim&\!\frac{\epsilon |\mathbf{w}_0|}{\rho_0}\sim\frac{2\pi}{\rho_0} \sqrt{\frac{Q_\eta (\eta-\eta_c) -\frac{\epsilon^2\langle\tilde{r}^2(\mathbf{X},T)\rangle}{6\rho_0^2}}{\mu}} \nonumber\\
\!&\sim&\! \frac{2\pi}{\rho_0} \sqrt{\frac{Q_1-1-\frac{\epsilon^2\langle\tilde{r}^2(\mathbf{X},T)\rangle}{6\rho_0^2}}{\mu}}. \label{eq71}
\end{eqnarray}

\subsection{Wavetrains and pulses}
Near the flocking transition, many authors have reported swarming in coherently moving high density bands separated by a low density gas \cite{gre04,cha08,sol15,hue08}. For lower values of the alignment noise, there is another transition to a polar liquid phase \cite{sol15}. Typically, these works carried out numerical simulations of the standard VM for small values of the average particle density $\rho_0$ and large box sizes. The band patterns appear as periodic wavetrain and as pulse solutions of the Toner-Tu equations \cite{cau14}. In the same vein, we can look for 1D traveling wave solutions of Eqs.~\eqref{eq55}-\eqref{eq56}, $r=r(\tau)$, $\mathbf{w}=(w(\tau),0)$, which depend on the moving coordinate $\tau=X+cT$. A more general assumption, $\mathbf{w}=|\mathbf{w}|\mathbf{e}_\Upsilon$ and $\tau=\hat{\mathbf{\kappa}}\cdot\mathbf{X}+cT$, inserted into Eq.~\eqref{eq56} simply produces $\hat{\mathbf{\kappa}}=\pm\mathbf{e}_\Upsilon$. This is compatible with our previous assumption if $\mathbf{e}_\Upsilon$ is parallel to the $X$ axis and we can select the sign of $c$. Eq.~\eqref{eq55} can be integrated once to yield
\begin{equation}
cr+\!\left(1+\frac{\epsilon r}{4\rho_0}\right)w=|\mathbf{w}_0|\Longrightarrow r=\frac{w_0-w}{c+\frac{\epsilon w}{4\rho_0}} .\label{eq72}
\end{equation}
Here we have chosen the integration constant so that it is compatible with the stationary solution given by Eq.~\eqref{eq58} and $r=0$. We have $w_0=|\mathbf{w}_0|=2\rho_0\sqrt{\eta_2Q_\eta/(M\gamma_0)}$, according to Eqs.~\eqref{eq58} and \eqref{a3_10}. Inserting Eq.~\eqref{eq72} into Eq.~\eqref{eq56} for a traveling wave, we obtain 
\begin{eqnarray}
&&\epsilon\delta \frac{d^2w}{d\tau}+F'(w)\frac{dw}{d\tau}+H'(w)=0,\label{eq73}\\
&&F(w)=\frac{w-w_0}{2\!\left(c+\frac{\epsilon w}{4\rho_0}\right)\!}\!\left(1+\frac{\epsilon(w-w_0)}{16\rho_0\!\left(c+\frac{\epsilon w}{4\rho_0}\right)\!}\right)\!-cw\nonumber\\
&&\quad\quad\quad+\frac{2+\gamma_0+(1-\gamma_0)M}{16\rho_0}\epsilon w^2, \label{eq74}\\
&&H'(w)= \frac{(w_0-w)w}{2\rho_0}\!\left[\!\left(1+\frac{\epsilon(w-w_0)}{3\!\left(c+\frac{\epsilon w}{4\rho_0}\right)\!}\right)\!\frac{1}{c+\frac{\epsilon w}{4\rho_0}}  \right.\quad\quad\nonumber\\
&&\quad\quad\quad\left.+\frac{\epsilon M\gamma_0(w+w_0)}{2\rho_0}\right]\!.\label{eq75}
\end{eqnarray}
Eq.~\eqref{eq73} has 
two constant solutions: $w=0$ and $w=w_0$. The eigenvalue problem determining their linear stability is 
\begin{eqnarray}
\lambda^2+\frac{F'(w)}{\epsilon\delta}\lambda+\frac{H''(w)}{\epsilon\delta}=0, \label{eq76}
\end{eqnarray}
where $w=0,\, w_0$. We have 
\begin{eqnarray}
H''(0)=\frac{w_0}{2\rho_0c}\!\left(1-\frac{\epsilon w_0}{3c}+\frac{\epsilon Mc\gamma_0w_0}{2\rho_0}\right)\!,\nonumber\\
H''(w_0)=-\frac{w_0}{2\rho_0}\!\left(\frac{1}{c+\frac{\epsilon w_0}{4\rho_0}}+\frac{\epsilon M\gamma_0w_0}{2\rho_0}\right)\!.\label{eq77}
\end{eqnarray}
Thus $H''(w_0)<0$ and $w=w_0$ is a saddle point, whereas the character of $w=0$ depends on the sign of the friction coefficient, $F'(0)=(2c)^{-1}-c+O(\epsilon)$. At $F'(0)=0$ or $c=c_0\sim 1/\sqrt{2}\approx 0.707$, there is a Hopf bifurcation and a stable limit cycle issues forth from the center $(0,0)$ [in the phase plane $(w,w'=dw/d\tau)$] for $c>c_0$. The limit cycle is a wavetrain of the amplitude Eqs.~\eqref{eq55}-\eqref{eq56}. Its amplitude increases with $c$ until it merges with two separatrices of the saddle point $(w_0,0)$ at a value $c=c_p>c_0$. The resulting saddle loop is a pulse of the amplitude Eqs.~\eqref{eq55}-\eqref{eq56}. Fig.~\ref{fig5} exhibits the particle densities of a wavetrain and a pulse that solve Eqs.~\eqref{eq72}-\eqref{eq73} for $c$ close to $c_p$. In a finite box with periodic boundary conditions, a pulse is a band of high particle density moving on a gas of low density that is recycled at the boundaries. A wavetrain appears as finitely many moving bands moving and recycling in the box, cf. Ref.~\cite{sol15}. 

\begin{figure}[h]
\begin{center}
\includegraphics[width=8cm]{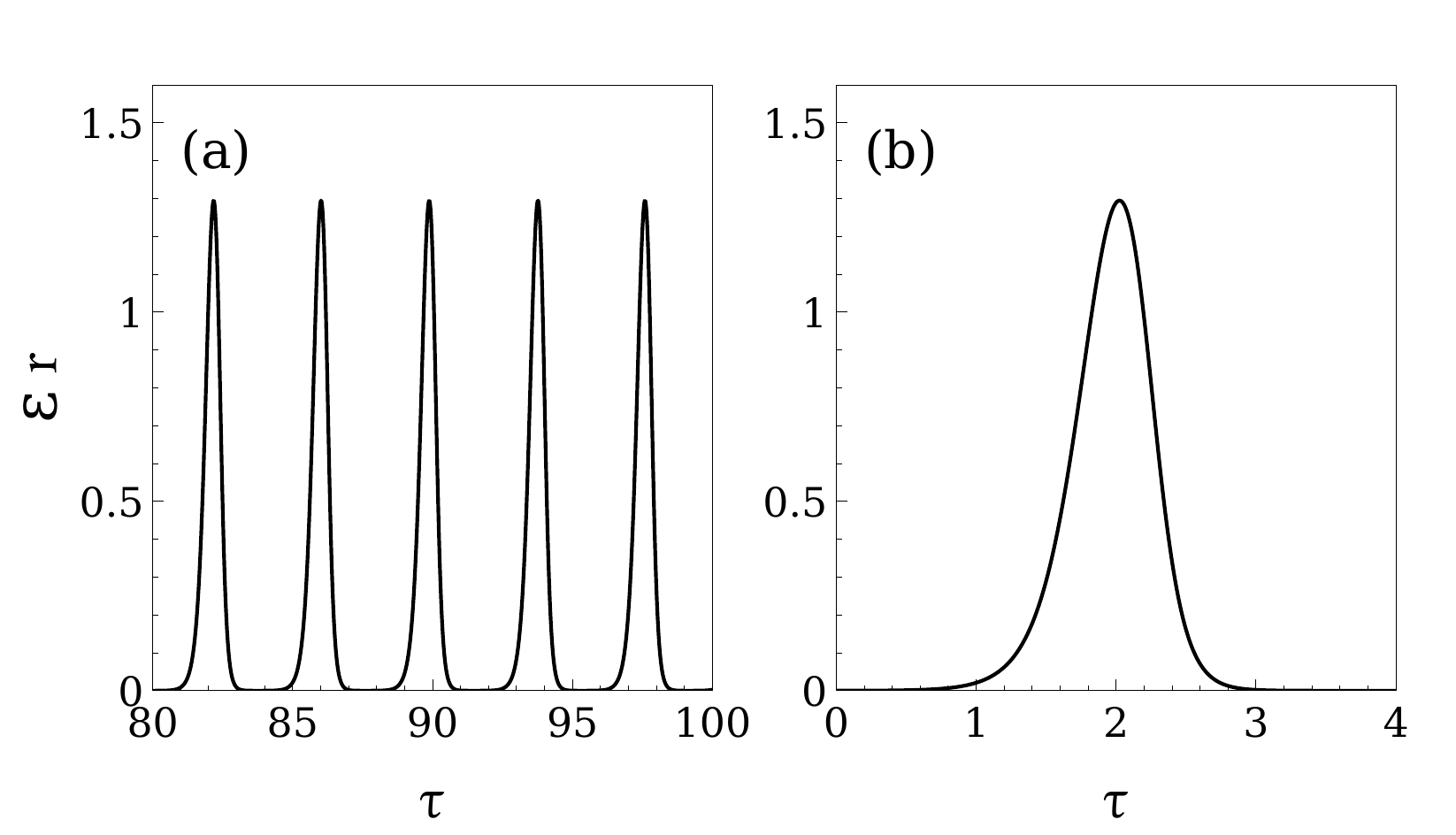}
\end{center}
\caption{Particle density $\epsilon r=\rho-\rho_0$ versus $\tau=X+cT$ for (a) wavetrain and (b) pulse solutions as trajectories of Eqs.~\eqref{eq72}-\eqref{eq73} with $\epsilon=0.1$, $M=7$, $\rho_0=10$, $c\approx 0.7125$.
 \label{fig5}}
\end{figure}

How do we find these traveling waves? The function $F(w)$ of Eq.~\eqref{eq74} has a local minimum at $w=w_1<w_0$ provided $c<c_M$ given by $F'(w_0)=0$:
\begin{eqnarray}\nonumber
c_M\!&=&\!\sqrt{\frac{1}{2}+\!\left[\frac{\epsilon Mw_0}{2\rho_0}\!\left(1-\gamma_0+\frac{4+\gamma_0}{4M}\right)\!\right]^2}\\
\!&+&\!\frac{\epsilon Mw_0}{2\rho_0}\!\left(1-\gamma_0+\frac{\gamma_0}{8M}\right)\!. \label{eq78}
\end{eqnarray}
In Eq.~\eqref{eq73}, $F'(w)<0$ amplifies oscillations for $w<w_1$ and $F'(w)>0$ damps them for $w_1<w<w_0$. As the limit cycle solution encircles $(w,w')=(0,0)$, the closer $w_1$ approaches $w_0$, the more `time' $\tau$ should $w(\tau)$ spend near $w_0$ for damping and amplification to compensate each other and produce a limit cycle. For $c>c_M$, there cannot be  limit cycle and homoclinic orbit solutions of Eq.~\eqref{eq73}, which are wavetrain and pulse solutions of Eqs.~\eqref{eq55}-\eqref{eq56}, respectively. Thus the existence and length of the interval $(c_0,c_p)$, with $c_p<c_M$, depends on the value of $\epsilon=|\eta-\eta_c|^{1/2}$ and the other parameters in Eq.~\eqref{eq73}. Since $c_M-c_0\sim\epsilon[1-\gamma_0-(2 \gamma_0-1)/(8M)] M w_0/(2\rho_0)$ as $\epsilon\to 0+$, the interval length $(c_p-c_0)$ is quite narrow (less than $c_M-c_0\approx 0.01894$ for the parameters in Fig.~\ref{fig5}). While the wavetrains and pulse of Fig.~\ref{fig5} move from right to the left, similar waves traveling from left to right can be similarly constructed replacing $w_0\to -w_0$, $c\to -c$.

In addition, the shapes of the wavetrains and the pulse depend on the value of $\delta$. As $\delta\to 0+$, the pulse decreases smoothly from $w=w_0$ (corresponding to the saddle point) up to a certain value $w=w_2$ ($w_2<w_1$), according to Eq.~\eqref{eq73} with $\delta=0$. Then it increases abruptly back to $w_0$ on the fast scale $\tilde{\tau}=(\tau-\tau_2)/(\epsilon\delta)$ according to the equation 
\begin{eqnarray}
\frac{d^2w}{d\tilde{\tau}^2}=-F'(w)\frac{dw}{d\tilde{\tau}}\Longrightarrow \frac{dw}{d\tilde{\tau}}= F(w_0)-F(w).\label{eq79}
\end{eqnarray} 
We have used that $w$ has to be $w_0$ and $dw/d\tilde{\tau}\to 0$ as $w$ approaches the saddle point. The condition that the solutions at the slow and fast scales have to match implies that   $w=w_2$ and $dw/d\tilde{\tau}=0$. Then we obtain the ``equal area rule'' $F(w_0)-F(w_2)=\int_{w_2}^{w_0}F'(w)dw=0$. One period of the wavetrain in the limit as $\delta\to 0+$ is described similarly. We start at a value $w=\tilde{w}_0<w_0$, which will fix the amplitude and period of the wavetrain. We solve Eq.~\eqref{eq73} with $\delta=0$ and initial condition $w=\tilde{w}_0$ until $w(\tau_2)=\tilde{w}_2>w_2$ such that $F(\tilde{w}_2)=F(\tilde{w}_0)$. Then $w$ increases abruptly back to $\tilde{w}_0$ according to Eq.~\eqref{eq79} with $\tilde{w}_0$ replacing $w_0$. Since $F'(w_1)=0$, the asymptotic description of the slow stage of wavetrains and pulses may involve additional internal layers. These  wavetrains and pulses with fast and slow stages are less symmetric than those displayed in Fig.~\ref{fig5}. 

 \section{Results of simulations of the Vicsek model}\label{sec:7}
We have performed numerical simulations of our modified VM in the different regions of Figure \ref{fig2} indicated by the linear stability analysis of the kinetic equation given in Section \ref{sec:3}. These regions and the predictions of bifurcation theory in Sections \ref{sec:5} and \ref{sec:6} are as follows: 
\begin{itemize}
\item[(I)] incoherent motion with polarization $W=0$; 
\item[(II)] stationary coherent motion with $W>0$ for $\xi_0=\pi$ and small $p$; 
\item[(III)] rotating wave phase with coherent motion (swarm rotation \cite{att14}), such that in Eq.~\eqref{eq4}, $W>0$, $\Upsilon=\Omega t+\Upsilon_0$ for $\xi_0<\pi$: the order parameter $Z(t)$ describes a circle about the origin at constant angular velocity as in Fig.~\ref{fig9}(a); see also Fig.~\ref{fig10}(a); 
\item[(IV)] coherent period 2 motion as in Figs.~\ref{fig9}(b) and Fig.~\ref{fig10}(b). 
\end{itemize}
At the critical lines separating the regions of Figs.~\ref{fig2}(a) and (b), we have found the following supercritical bifurcations from incoherent motion with uniform particle density: pitchfork (I to II), Hopf (I-III), and period doubling (I-IV). 

\begin{figure}[h]
\begin{center}
\includegraphics[width=8cm]{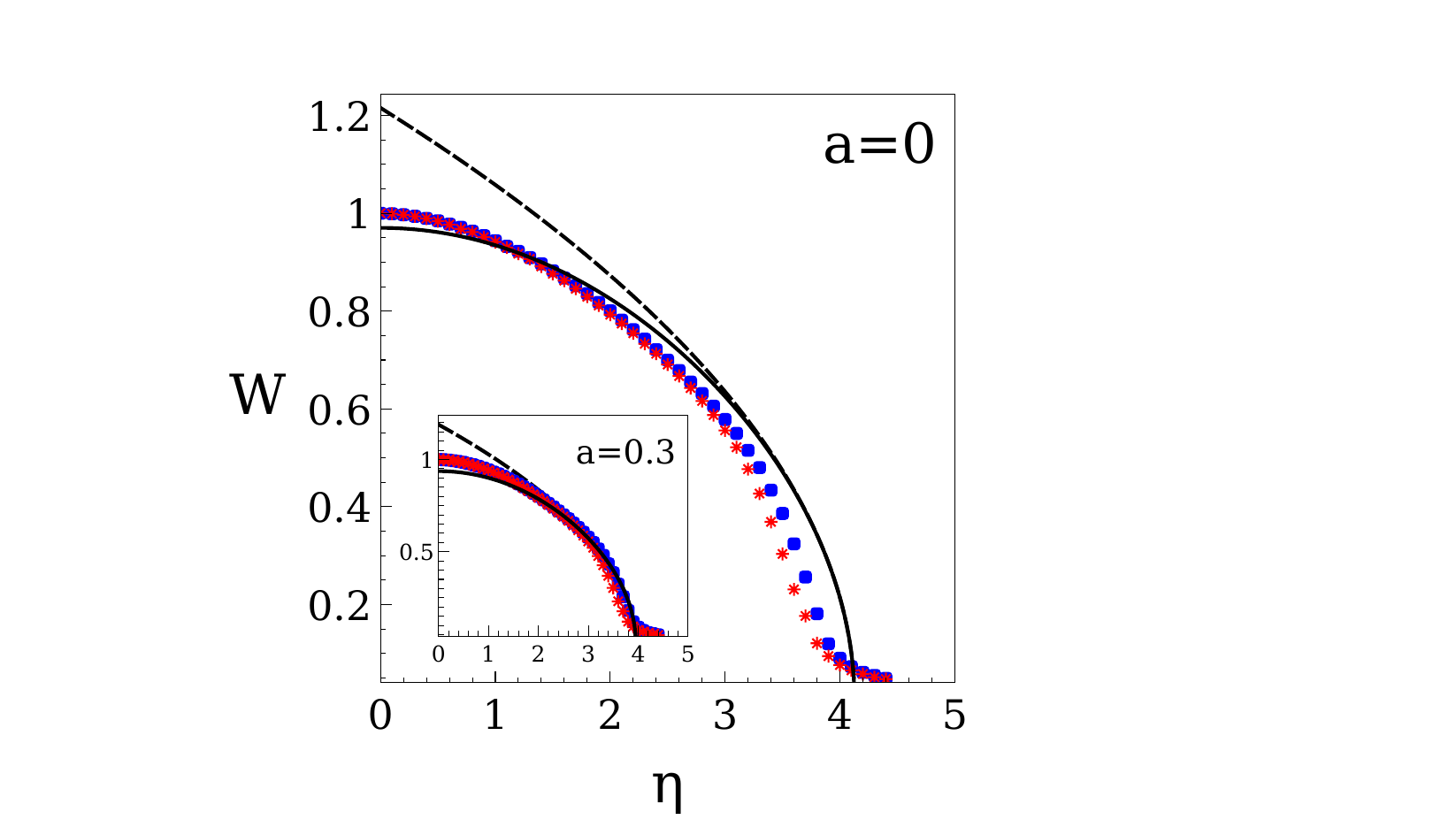}
\end{center}
\caption{Polarization $W$ versus $\eta$ for $p=0$ (standard VM), $M=7$, $N=1000$, and $\rho_0=10$ (blue squares) or $\rho_0=5$ (red asterisks). Dashed and solid lines correspond to Eqs.~\eqref{eq59} and \eqref{eq60}, respectively. Inset: Same graph with the critical noise shifted according to Eq.~\eqref{eq71} with $a=\epsilon\sqrt{\langle \tilde{r}^2\rangle/6}/\rho_0=0.3$.
 \label{fig6}}
\end{figure}

\begin{figure}[h]
\begin{center}
\includegraphics[width=8.5cm]{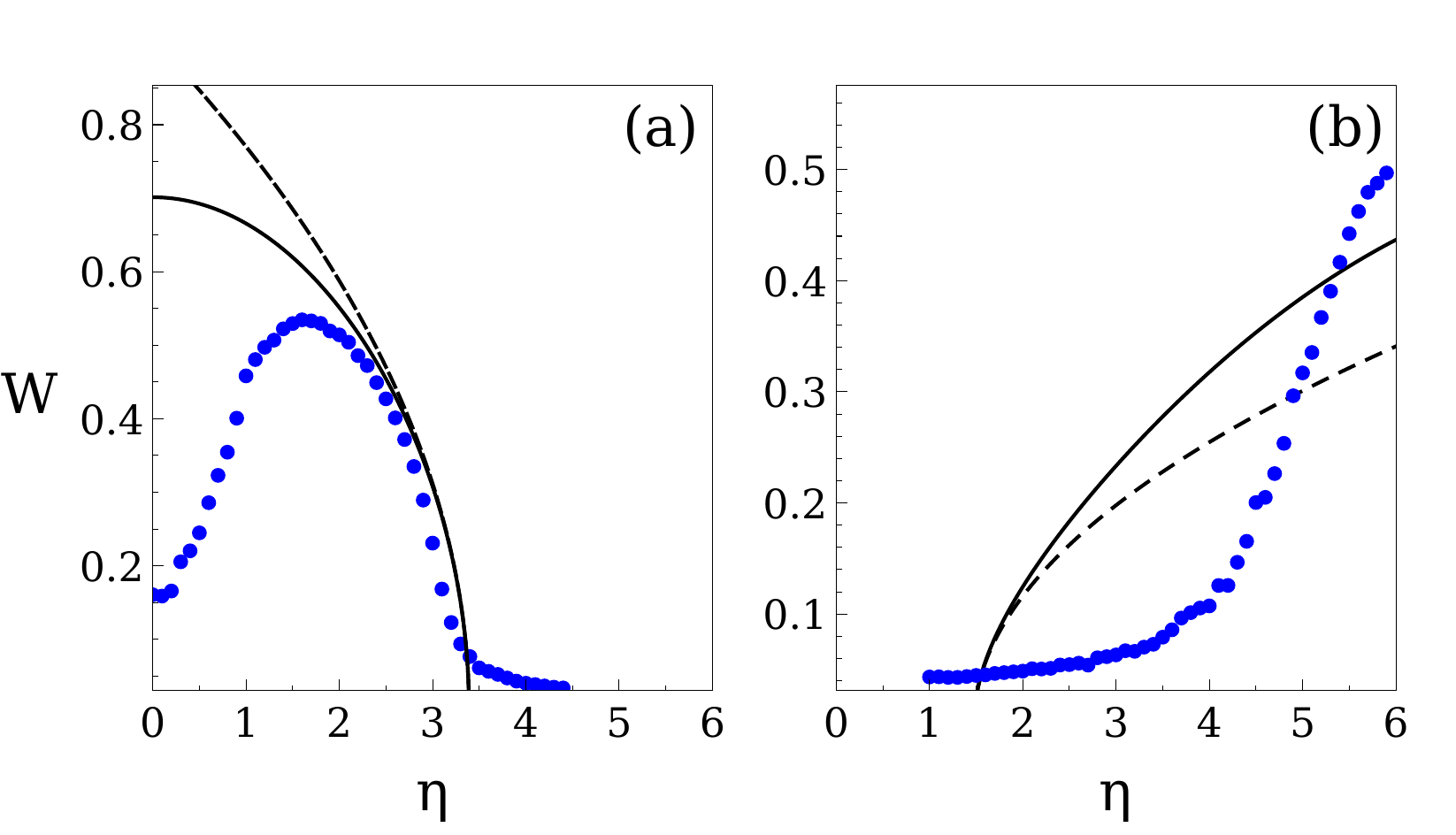}
\end{center}
\caption{Polarization $W$ versus $\eta$ for $\xi_0=\pi$, $M=7$, $\rho_0=10$, $N=1000$, and (a)  lower sector in Fig.~\ref{fig2}(b) ($p=0.1$); (b) upper sector in Fig.~\ref{fig2}(b) ($p=0.7$). Circles are numerical simulation data whereas solid and dashed lines correspond to kinetic theory approximations of Eqs.~\eqref{eq60} or \eqref{eq59}, respectively. In (b), flocking is heterogeneous for any $\eta$ and results from simulations are quite different from theoretical lines corresponding to the uniform amplitude. 
 \label{fig7}}
\end{figure}

\begin{figure}[h]
\begin{center}
\includegraphics[width=5cm]{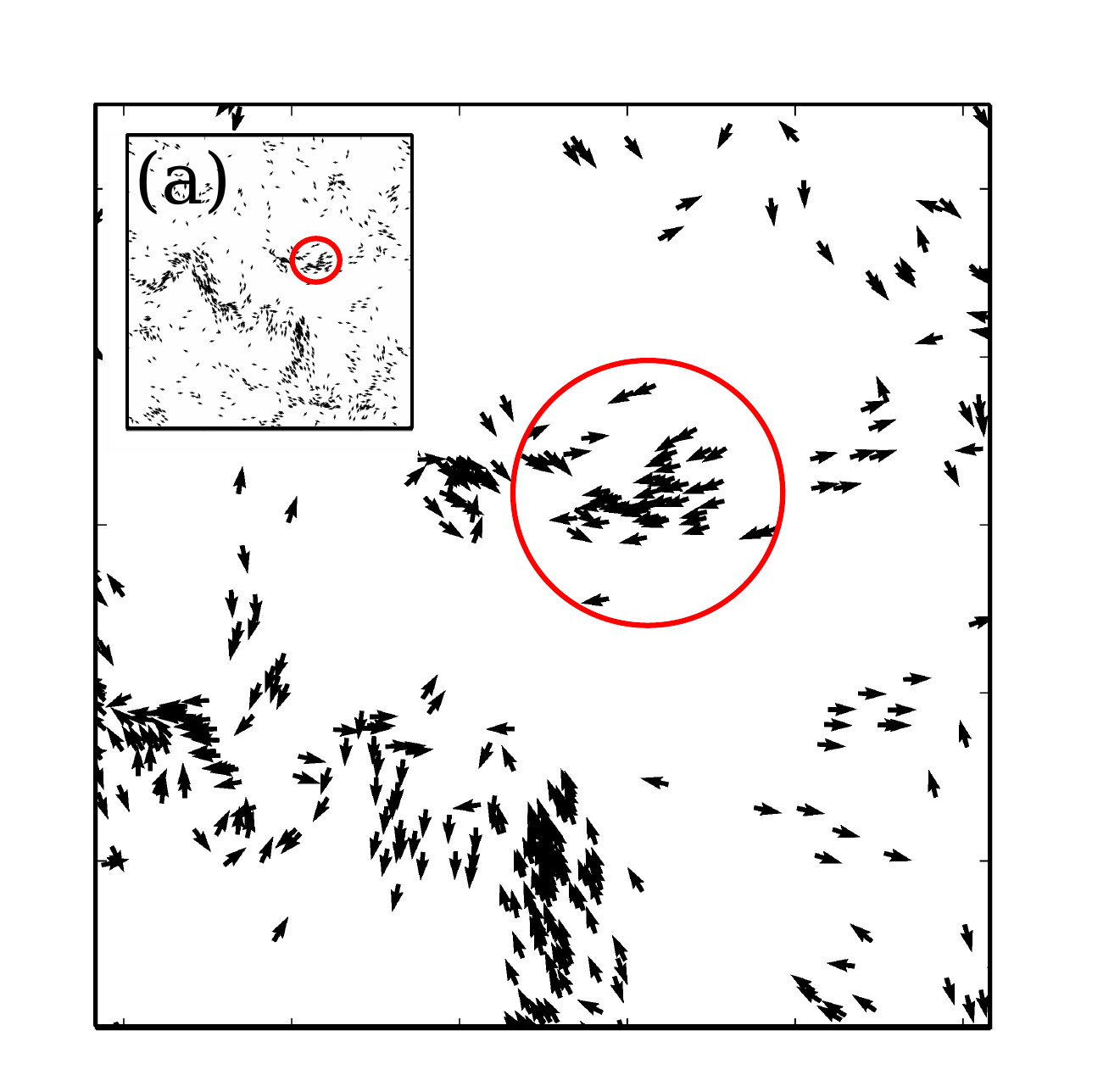}
\includegraphics[width=5cm]{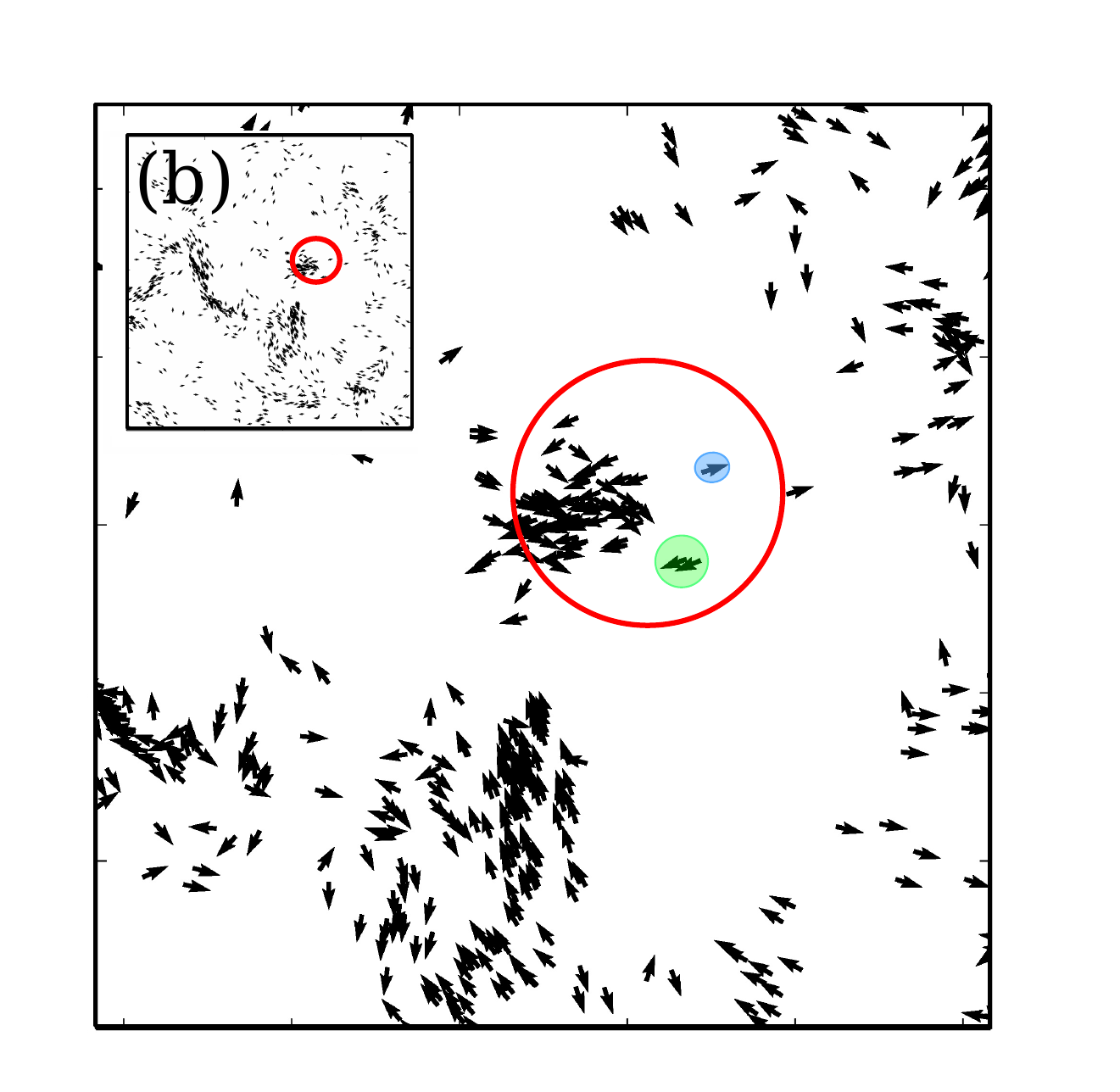}
\includegraphics[width=5cm]{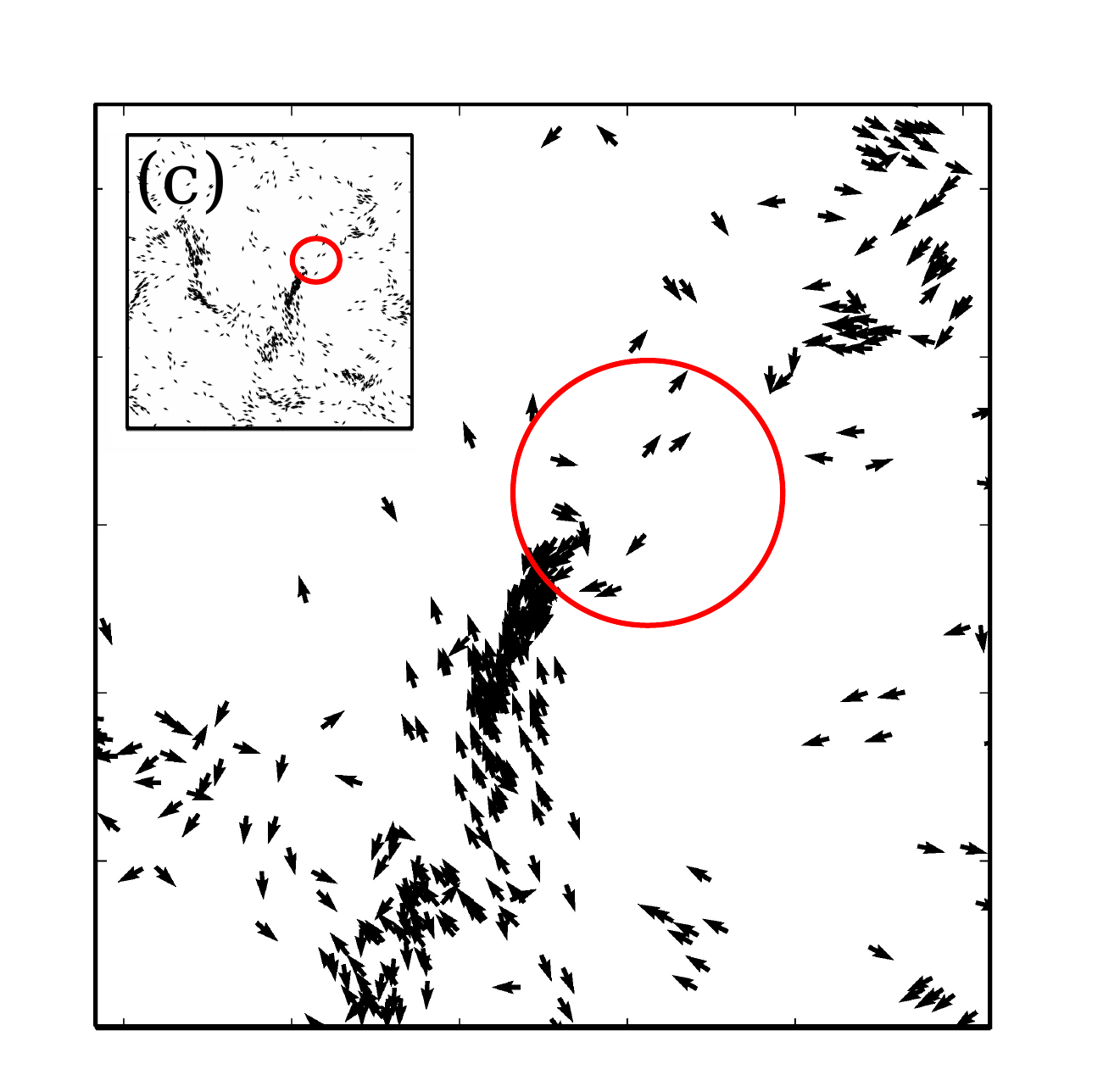}
\end{center}
\caption{Three snapshots (a), (b) and (c) showing heterogeneous clustering for particles with contrarian compulsions. The insets show the position of the red circle on the complete box. A cluster of particles leaves the red circle, joins another set of particles and forms an elongated cluster after some time. Close-by particles may have one direction and its contrary, as the marked particles in panel (b). Parameter values are $\xi_0=\pi$, $M=7$, $\rho_0=3$, $p=\eta=0.1$, $N=1000$. 
 \label{fig8}}
\end{figure}

\paragraph{Pitchfork bifurcation.} Figs.~\ref{fig6} and \ref{fig7}(a) compare the polarization obtained from direct numerical simulations of the VM with the theoretical curves of Eqs.~\eqref{eq59} and \eqref{eq60} for $p=0$ and $p=0.1$ (transition I-II in Fig.~\ref{fig2}), respectively. The polarization shown in these figures is an ensemble average over 10 replicas of the stochastic process. In Fig.~\ref{fig6}, we observe that the results of numerical simulations tend to the uniform solution predicted by bifurcation theory as we increase the density $\rho_0$. This indicates that the simulations produce solutions that are closer to be independent of space. The shift in the bifurcation point $\eta_c$ observed in simulations may be corrected when we approximate better $Q_1$ in Eq.~\eqref{eq20} and take into account the shift given by Eq.~\eqref{eq71}; see Ref.~\cite{BT18}. However in Fig.~\ref{fig7}(a), the results of simulations depart appreciably from the theoretical prediction, which indicates that flocking is not homogeneous. Although in both cases the bifurcation is pitchfork, there is a striking difference between the polarization curves for the standard and modified VM. For the standard VM, the maximum polarization ($W=1$) is reached for zero alignment noise $\eta=0$. However, for $p=0.1$ [transition I-II in Fig.~\ref{fig2}(b)], the polarization shown in Fig.~\ref{fig7}(a) is maximal for a nonzero value of $\eta$. A similar behavior also occurs for RWPs, as explained later in this section. Fig.~\ref{fig8} shows three snapshots of 1000 particles for the VM with contrarian compulsions for $M=7$, $\rho_0=3$, $p=\eta=0.1$. The insets show the location of all the particles in three time instants. We observe that there are small clusters that move and persist in time, with dynamics as shown in Fig.~\ref{fig8}. Had different realizations of the stochastic process shown one large cluster and a number of freely moving particles, the result of ensemble averaging would be close to that of Eq.~\eqref{eq60} for homogeneous flocking. Ensemble averages of heterogeneous flocking such as that depicted in Fig.~\ref{fig8} give a polarization that differs markedly from Eq.~\eqref{eq60}, as shown in Fig.~\ref{fig7}(a).

\begin{figure}[h]
\begin{center}
\includegraphics[width=8.5cm]{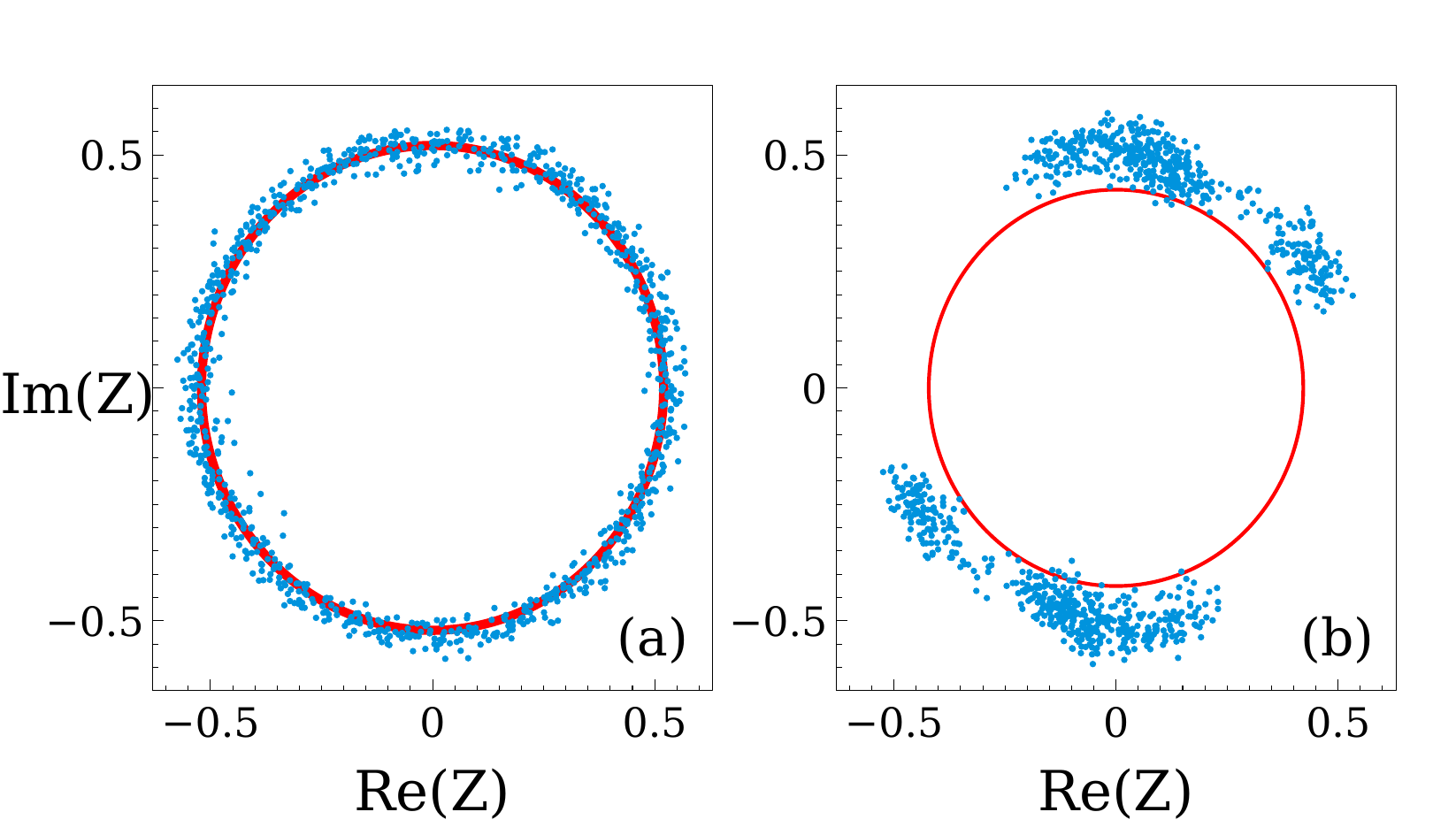}
\end{center}
\caption{Complex flocking order parameter $Z(t)$ for $M=7$, $\rho_0=10$, $N=1000$, and: (a) $\xi_0=3\pi/4$, $p=0.1$, $\eta=2.5$, marked by an arrow in Figure \ref{fig11}(a). The order parameter for the RWP traces a circle (theory: solid line), implying a nonzero angular velocity. (b) $\xi_0=\pi$, $p=0.7$, $\eta=5.9$. The order parameter alternates between points with a phase difference of $\pi$, marking abrupt oscillations in flocking (theory: two  opposite points on the circle). Sampling time is 1 and transients have been eliminated.  \label{fig9}}
\end{figure}

\begin{figure}[h]
\begin{center}
\includegraphics[width=8.5cm]{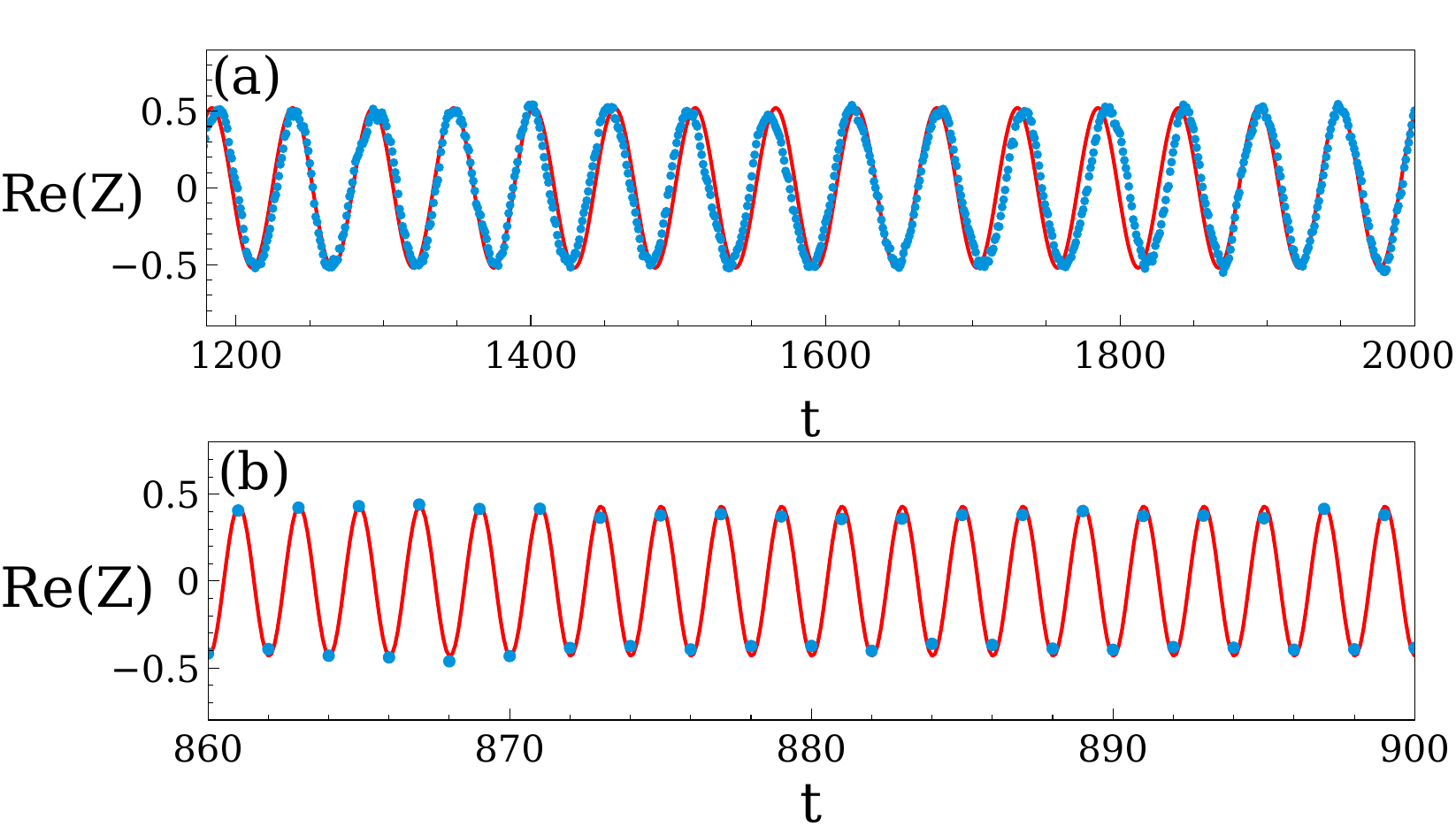}
\end{center}
\caption{Real part of $Z(t)$ corresponding to: (a) Figure \ref{fig9}(a); and (b) Fig.~\ref{fig9}(b). Numerical simulations (circles) are compared to spatially uniform phases from kinetic theory (solid lines) given by Eq.~\eqref{eq45} with $\psi=0$. }
 \label{fig10}
\end{figure}

\begin{figure}[h]
\begin{center}
\includegraphics[width=8.5cm]{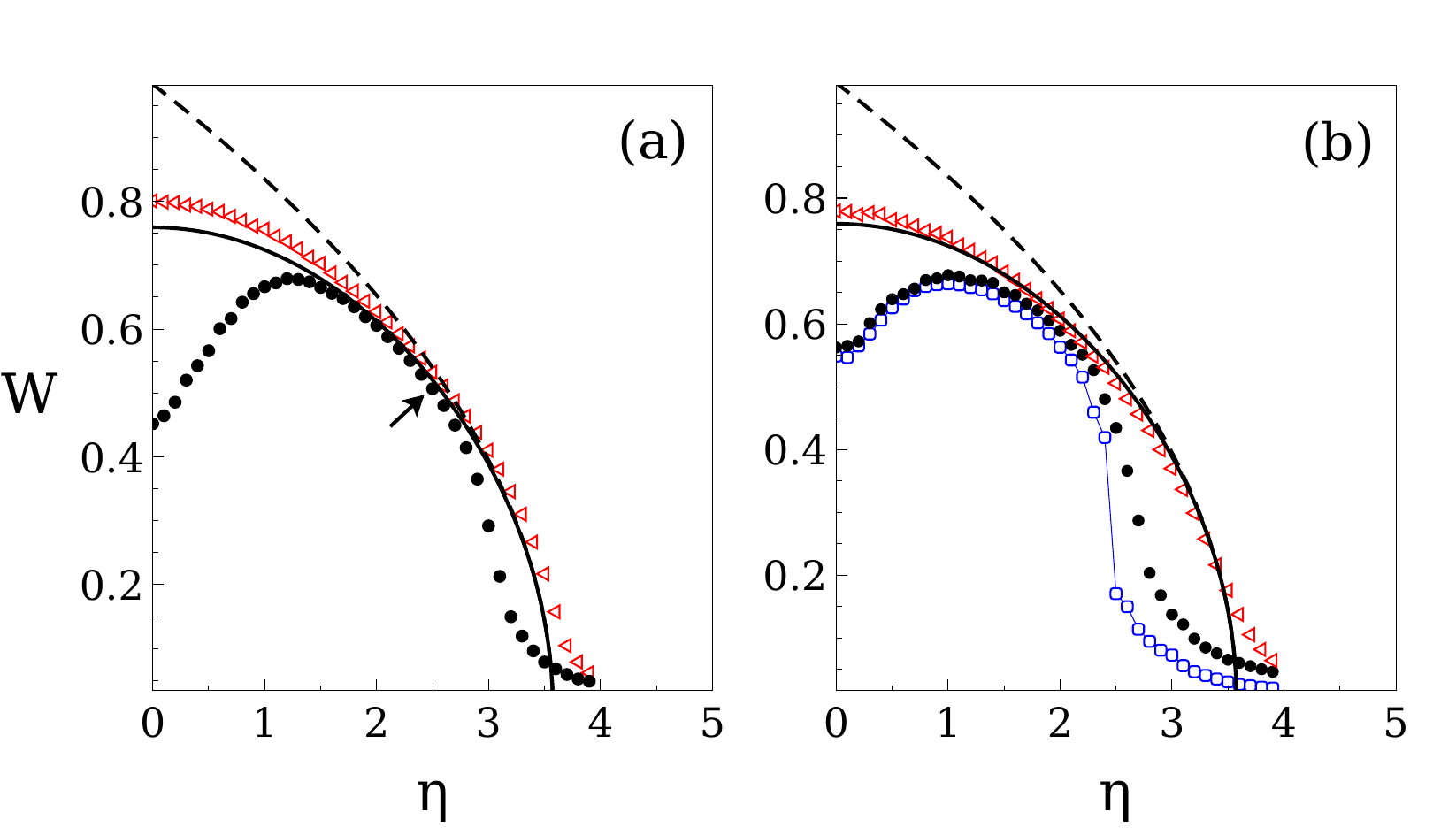}
\end{center} 
\caption{Polarization $W$ versus $\eta$ for the lower sector III of Fig.~\ref{fig2}(a). Simulation data obtained by ensemble averages over 10 replicas of the VM for $\xi_0=3\pi/4$, $p=0.1$, $M=7$, and (a) $\rho_0=10$, $N=1000$; (b) $\rho_0=3$, $N=1000$ (filled circles) and $N=5000$ (blue circles). Empty red triangles arise from simulating the VM with backward update instead of Eq.~\eqref{eq2}. Solid and dashed lines correspond to kinetic theory approximations of Eqs.~\eqref{eq45} and \eqref{eq44}, respectively. As $\rho_0$ increases, simulation data approach the line of the spatially uniform RWP with order parameter given by Eq.~\eqref{eq45}. Departure from this line indicates a nonuniform RWP. Near $\eta_c$,  simulation data depart from the theoretical curve due to finite size effects. Panel (b) shows that the transition I-III becomes discontinuous for Eqs.~\eqref{eq1}-\eqref{eq2} at larger box sizes ($L=\sqrt{N/\rho_0}$).  \label{fig11}}
\end{figure}

\begin{figure}[h]
\begin{center}
\includegraphics[width=5cm]{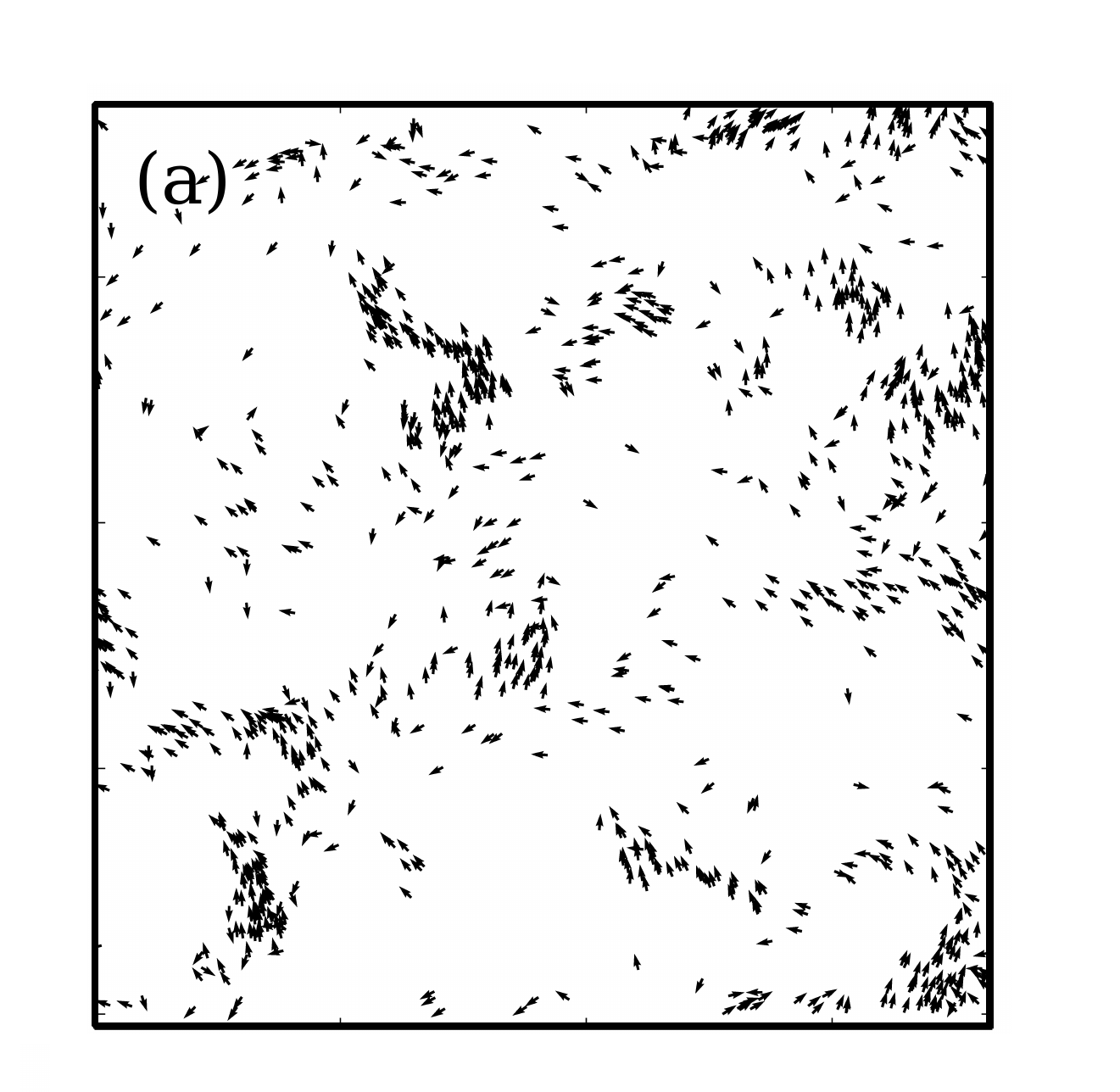}
\includegraphics[width=5cm]{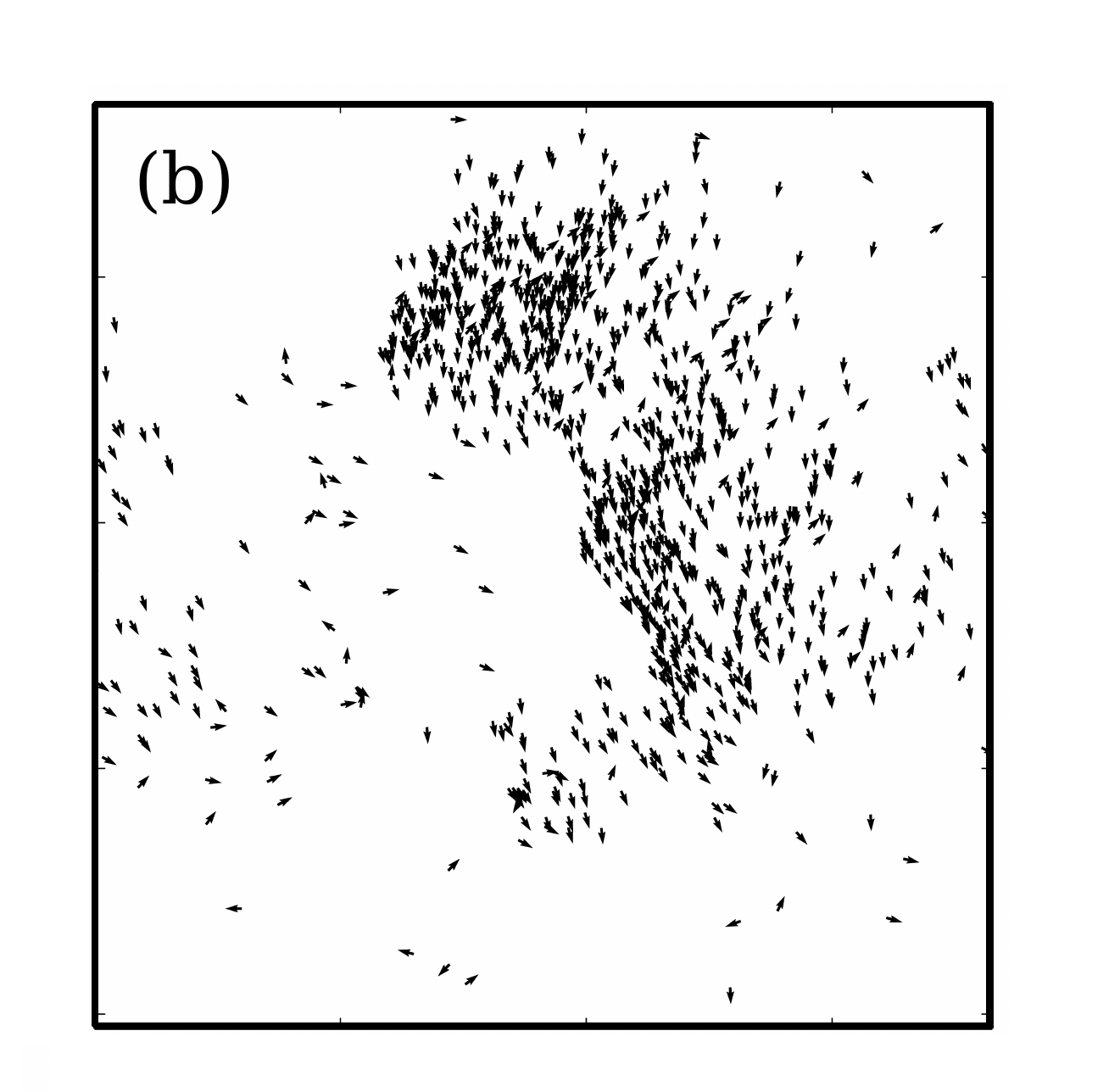}
\end{center}
\caption{Two snapshots showing clustering for particles with almost contrarian compulsions for (a) forward, and (b) backward update. With forward update, clustering is heterogeneous, whereas with backward update, nonzero polarization is attained by one large cluster and a gas of particles outside it. Parameter values are $\xi_0=3\pi/4$, $M=7$, $\rho_0=3$, $p=\eta=0.1$, $N=1000$. Values of the polarization $W$ are: (a) 0.57, (b) 0.78. 
 \label{fig12}}
\end{figure}

\paragraph{Hopf bifurcation.} Figs.~\ref{fig9}(a) and \ref{fig10}(a) show that the complex order parameter of the RWP in the lower region III of Fig.~\ref{fig2}(a) is close to the uniform values of Eqs.~\eqref{eq44}-\eqref{eq45} predicted by bifurcation theory. The agreement between simulations of the VM and uniform solutions predicted by bifurcation theory is not as good for the period 2 phase, as shown in Figs.~\ref{fig9}(b) and \ref{fig10}(b). Figure \ref{fig11} describes the transition I-III in the lower sector of Fig.~\ref{fig2}(a) for almost contrarian compulsions. Solid and dashed lines correspond to spatially uniform coherent phases calculated from kinetic theory. Departure of simulation data from these lines indicates heterogeneous RWPs. Observe the dispersion of simulation data in Fig.~\ref{fig9}(a). As the average density $\rho_0=N/L^2$ increases, the phases become more uniform and simulation data approach theoretical predictions. It is interesting to note that simulations of the VM with backward update produce polarizations closest to the theoretical curves except very close to the bifurcation point (which is due to finite size effects). Fig.~\ref{fig12}(a) shows that, for small values of $p$ and $\eta$, forward update with almost contrarian compulsions produces small clusters and many seemingly free particles. Compared with the same VM and parameters but with backward update, Fig.~\ref{fig12}(b) shows one large cluster and a small number of free particles. When averaging over many replicas of the stochastic process, backward update produces a polarization consistent with the homogeneous flocking prediction of Eq.~\eqref{eq45}. However, the persistence of several clusters in the case of forward update produces a polarization that does not correspond to homogeneous flocking.

For the VM of Eqs.~\eqref{eq1}-\eqref{eq3} with forward update, there is an appreciable shift of the simulation data to smaller values of the noise $\eta$ near the bifurcation point. This shift decreases as the average density $\rho_0$ increases but it does not disappear, as the comparison between Figs.~\ref{fig11}(a) and \ref{fig11}(b) indicates. The same shift also occurs in the standard VM: Even for an average density as large as $\rho_0=20.4$, the critical noise $\eta_c$ is  different in direct simulations and theory, as noted in the different rescaling in Fig.~3 of Ref.~\cite{cho12}. Fig.~\ref{fig11}(b) shows that $W$ departs the solid line and the bifurcation becomes discontinuous for box size beyond a critical value. This change also occurs in the standard VM, except that the phase transition is from incoherence to flocking with nonzero average velocity \cite{gre04,cha08,sol15,bag09}. Fig.~10 of Ref.~\cite{cha08} shows that the critical length decreases as the average particle density increases for the standard VM. The density $\rho_0=3$ considered in Fig.~\ref{fig11}(b) is three times larger than the largest one in Fig.~10 of Ref.~\cite{cha08} ($\rho_0=1$ in our nondimensional units), which is why we observe a discontinuous transition for a length as small as $L=\sqrt{5000/3}=40.8$.

\paragraph{Period doubling bifurcation.} Strikingly, when the probability of contrarian movement is sufficiently high, increasing the alignment noise favors time periodic, spatially heterogeneous, flocking. Increasing $\eta$ (tolerance to failure in particle alignment) in the upper part of Region I in Fig.~\ref{fig2} favors forming clusters. For $p>p_0\approx 2/\sqrt{\pi M}$ [cf. Eq.~\eqref{eq20}] and small $\eta$, preponderance of contrarian over conformist motion ensures incoherence of motion. The upper lines in Figure \ref{fig2} suggest transition from incoherent motion to a degree of flocking as $\eta$ surpasses a critical value. Figures \ref{fig7}(b) and \ref{fig9}(b) for contrarian compulsions show that the polarization $W$ calculated from simulations departs markedly from the theoretical line corresponding to uniform density. Figures \ref{fig9}(b) and \ref{fig10}(b) illustrate that the complex order parameter alternates between numbers with arguments differing by $\pi$. Thus, the ordered phase is periodic in time with period 2 although, as shown in Figs.~\ref{fig9}(b) and \ref{fig10}(b), the amplitude of the oscillation has an envelope that wanders in a certain region of the complex plane. The dispersion of points in Figure \ref{fig9}(b) is due to the formation of clusters with varying size that change in time and produce a nonuniform density. The persistence of heterogeneous clustering yields ensemble averaged polarizations that differ from the prediction of Eq.~\eqref{eq45} with $\Omega=\pi$ for spatially homogeneous flocking. Sufficient tolerance to failures in the alignment of the conformist particles keeps nonuniform flocking at the expense of back and forth motion of the flocks between opposite average phases of the order parameter. Active particles in exotic phases perform rotations and oscillations, not just translations as in the standard VM; see the movies in the Supplementary Material \cite{suppl}. 

\paragraph{Optimal noise.} Figures \ref{fig7}(a) and \ref{fig11} indicate that the polarization $W$ increases with alignment noise until it reaches a maximum: {\em To attain maximum flocking we need an optimum degree of alignment noise}, for both Hopf and pitchfork bifurcations in the lower sectors of Fig.~\ref{fig2}(a) and (b). For larger noise, $W$ decreases and the values obtained from simulations approach the theoretical curve for uniform particle density. With low probability of contrarian motion and small $\eta$, particle clusters  form, move coherently and change size (heterogeneous flocking), as shown in Fig.~\ref{fig8} and \ref{fig12}(a). See also the movies in Ref.~\cite{suppl}.

\section{Final Remarks}\label{sec:8}
In conclusion, we have proposed a modified Vicsek model in which active particles may align their velocity with the local average direction of motion or with the (almost) opposite direction. Flocking behavior depends on the probability $p$ of (almost) contrarian compulsions compared with that of conformist alignment according to the Vicsek rule. From incoherent motion with negligible polarization, we have found far from equilibrium transitions to ordered stationary, rotating wave, and period 2 phases. According to our analysis of the VM kinetic theory, exotic time dependent phases issue forth from uniform incoherent motion as period doubling or Hopf bifurcations described by real or complex Ginzburg-Landau equations, respectively. Departure of simulation data from spatially uniform states of the above types corresponds to similar nonuniform phases. Strikingly, increasing the alignment noise $\eta$ may favor order. For small $p$, there is a nonzero optimal value of $\eta$ that achieves maximum polarization. When $p$ is close to 1, increasing $\eta$ may transform incoherent particle motion into a coherent ordered phase.

Stable phases with time dependent order parameter may have appeared in earlier work. For example, Chat\'e {\em et al} introduced a Vicsek-like model for apolar nematic active particles can move with equal probability along their orientation $\theta_i$ or along the contrarian orientation $\theta_i+\pi$ \cite{cha06}. At one time step later, the orientation is chosen as the first eigenvector of a tensor order parameter plus disorder noise. They observed a continuous transition similar to the Kosterlitz-Thouless transition, characterized by large spatial fluctuations of the time-averaged order parameter \cite{cha06}. This is different from the discontinuous transition observed for the standard (polar) VM if the box size is sufficiently large \cite{gre04}. In the case of our modified VM with contrarian or almost contrarian compulsions, the continuous bifurcation becomes discontinuous for sufficiently large box sizes. Liebchen and Levis proposed a continuous-time model consisting of Langevin equations for the angles $\theta_i$ of the particle polarizations. In the model, the angular velocities $\dot{\theta}_i$ equal a constant rotation $\omega$, plus a Kuramoto coupling to the angles of particles inside the circle of influence of particle $i$, plus i.i.d. white noise terms $\xi_i(t)$ \cite{lie17}. The model presents a flocking transition from a disordered gas phase to one or several clusters comprising particles rotating in synchrony (microflocking). Clustering in our VM with almost contrarian compulsions is reminiscent of microflocking in Ref.~\cite{lie17}. In our case, the probability of deflecting a large angle from the conformist mean direction is responsible for the rotation inside heterogeneous flocks, and we do not need to impose an external common angular velocity to achieve rotating clusters. Menzel \cite{men12} studied a similar model to that in Ref.~\cite{lie17} for two different populations of particles, that had Kuramoto coupling but without the constant rotation. He found a variety of behaviors including clustering and stripe patterns, but not clusters of synchronously rotating particles. Lastly, Chepizhko {\em et al} considered a similar model without constant rotation and for a single particle species, which interacts with obstacles that could be fixed or diffusing in space, \cite{che13}. For this quite different system, they observed that the time-averaged order parameter exhibited a maximum for an optimal noise strength, a phenomenon similar to that presented in our Figs.~\ref{fig7}(a) and \ref{fig11}.

In future work, it will be interesting to see whether direct simulations of the VM produce patterns with local order similar to those found for the CGLE. While the predictions from our model could be applicable to social systems (e.g., opinion formation models \cite{heg02,kur16}, emergency escape of a crowd from a confined region with several exits \cite{hag17}), they may be tested experimentally by devising appropriate robot swarms; see \cite{rub14}.

\acknowledgements
We thank Antonio Lasanta for useful comments and for bringing the kinetic theory work of Ihle and collaborators to our attention. This work has been supported by the Ministerio de Econom\'\i a y Competitividad grants  MTM2014-56948-C2-2-P and MTM2017-84446-C2-2-R. LLB thanks Russel Caflisch for hospitality during a sabbatical stay at the Courant Institute and acknowledges support of the Ministerio de Ciencia, Innovaci\'on y Universidades ``Salvador de Madariaga'' grant PRX18/00166.

\appendix 
\section{Definition of the Vicsek model in dimensional units}\label{ap:1}
We consider an angular noise Vicsek model with forward updating rule. Our choice differs from Vicsek et al.'s \cite{vic95} in the updating rule and it is the same as in \cite{ihl16}. See Refs.~\cite{hue08,bag09} for a discussion on how different definitions of the VM affect the character of the order-disorder phase transition. 

More specifically, in dimensional units, $N$ particles with positions $\mathbf{x}_j$ and velocities $\mathbf{v}_j=v_0(\cos\theta_j,\sin\theta_j)$, $j=1,\ldots,N$, are inside a square box of size $L$ and we use periodic boundary conditions. The particles undergo discrete dynamics so that their positions are forwardly updated, 
\begin{eqnarray}
\mathbf{x}_j(t+\Delta t)=\mathbf{x}_j(t)+ \mathbf{v}_j(t+\Delta t)\,\Delta t.\label{s1}
\end{eqnarray}
Here $t=0,\Delta t, 2\Delta t,\ldots$. The angle of a particle $i$ is updated according to the Vicsek angular noise rule 
\begin{eqnarray}
\theta_i(t+\Delta t)=\mbox{Arg}\left(\sum_{|\mathbf{x}_j-\mathbf{x}_i|<R_0}e^{i\theta_j(t)}\right)+\xi_i(t),\label{s2}
\end{eqnarray}
where we sum over all particles that, at time $t$, are inside a circle of radius $R_0$ centered at $\mathbf{x}_i$. The sum includes the particle $i$. At each time, $\xi_i(t)$ is a random number chosen with probability density $g(\xi)$ as indicated by Eq. (\ref{eq3}). 

We nondimensionalize the model by measuring velocity in units ov $v_0$, time in units of $\Delta t$ and space and lengths in units of $v_0\Delta t$. In these units, $v_0=1$, $\Delta t=1$, and the nondimensional average particle density becomes
\begin{eqnarray}
\rho_0=\frac{Nv_0^2(\Delta t)^2}{L^2}, \label{s3}
\end{eqnarray}
whereas the average number of neighbors of a particle, $M=N\pi R_0^2/L^2$, remains an unchanged dimensionless parameter.

\section{Methods} \label{ap:2}
\paragraph{Numerical methods.} At each time step, we have updated the position and velocity of $N$ active particles using Eqs.~\eqref{eq1}-\eqref{eq2}. To select the values $\xi_i$, we first choose a number between 0 and 1 randomly with uniform probability. If this number is $p$ or smaller, we set $\xi=\xi_0$. Otherwise, $\xi$ is selected randomly in the interval $(-\eta/2,\eta/2)$.

\paragraph{Regular perturbation theory for the eigenvalues of $\mathcal{M}_{ij}$.}
Assume that the matrix $\mathcal{M}(Q)=\mathcal{D}(Q)+\varepsilon\mathcal{N}(Q)$, where $\varepsilon\mathcal{N}_{ij}(Q)$ are the off-diagonal terms of $\mathcal{M}(Q)$ and $\varepsilon\ll 1$. The eigenvalues, $\lambda(\varepsilon)=\sum_{j=0}^\infty \lambda^{(j)}\varepsilon^j$, and eigenfunctions $\psi(\varepsilon)=\sum_{j=0}^\infty \psi^{(j)}\varepsilon^j$ of $\mathcal{M}(Q)$ can be expanded in powers of the scaling parameter $\varepsilon$ and inserted in the eigenvalue equation for $\mathcal{M}(Q)$. As $\varepsilon\to 0$, we obtain the hierarchy of linear equations
\begin{eqnarray}
&&(\mathcal{D}-\lambda^{(0)}\mathcal{I})\psi^{(0)}=0,\label{a2_1}\\
&&(\mathcal{D}-\lambda^{(0)}\mathcal{I})\psi^{(1)}=(\lambda^{(1)}\mathcal{I}-\mathcal{N})\psi^{(0)},\label{a2_2}\\
&&(\mathcal{D}-\lambda^{(0)}\mathcal{I})\psi^{(2)}=(\lambda^{(1)}\mathcal{I}-\mathcal{N})\psi^{(1)}+\lambda^{(2)}\psi^{(0)}, \quad\quad\label{a2_3}
\end{eqnarray}
etc. The first equation says that $\lambda^{(0)}$ are the diagonal elements of $\mathcal{M}$, $\lambda^{(0)}_j=QJ_0(|\mathbf{K}|)-C^{(1)}[e^{ij\theta}]_j$ ($j=1,\ldots$). For the other non-homogeneous linear equations to have solutions, their right hand sides have to be orthogonal to the eigenvectors $\psi^{(0)}$. The $j$th eigenvector has components $\delta_{lj}$. The orthogonality condition for Eq.~\eqref{a2_1} produces $\lambda^{(1)}_j=\mathcal{N}_{jj}=0$. Eq.~\eqref{a2_2} becomes 
\begin{eqnarray}
(\lambda^{(0)}_i-\lambda^{(0)}_j)\psi^{(1)}_i=-\mathcal{N}_{ij}\Longrightarrow
\psi^{(1)}_i=\frac{\mathcal{N}_{ij}}{\lambda^{(0)}_j-\lambda^{(0)}_i}, \label{a2_4}
\end{eqnarray}
for $i\neq j$ and $\psi^{(1)}_j=0$. We now insert Eq.~\eqref{a2_4} into Eq.~\eqref{a2_3} and use the orthogonality condition to obtain
\begin{eqnarray}
&&\varepsilon^2\lambda^{(2)}_j=\varepsilon^2\sum_{l, l\neq j}\frac{\mathcal{N}_{jl}\,\mathcal{N}_{lj}}{\lambda^{(0)}_j-\lambda^{(0)}_l}\nonumber\\
&&\quad= -Q^2\sum_{l, l\neq j}\frac{(e^{i\mathbf{K}\cdot\mathbf{v}+i(j-l)\theta})_0(e^{i\mathbf{K}\cdot\mathbf{v}+i(l-j)\theta})_0}{C^{(1)}[e^{ij\theta}]_j-C^{(1)}[e^{il\theta}]_l},\quad \label{a2_5}
\end{eqnarray}
with $\mathbf{v}=(\cos\theta,\sin\theta)$. In this expression, we have factors:
\begin{eqnarray}
(e^{i\mathbf{K}\cdot\mathbf{v}+in\theta})_0 \!&=&\! \frac{1}{2\pi}\int_{-\pi}^\pi e^{i|\mathbf{K}|\cos(\theta-\mbox{\footnotesize{Arg}}\mathbf{K})+in\theta}d\theta\nonumber\\
\!&=&\! \frac{e^{in(\mbox{\footnotesize{Arg}}\mathbf{K}+\pi/2)}}{2\pi}\int_{-\pi}^\pi e^{i|\mathbf{K}|\sin\zeta-in\zeta}d\zeta\nonumber\\
\!&=&\! e^{in(\mbox{\footnotesize{Arg}}\mathbf{K}+\pi/2)}J_n(|\mathbf{K}|),\label{a2_6}
\end{eqnarray}
where we have changed variable $\theta=$Arg$\mathbf{K}+\pi/2-\zeta$, shifted the limits of integration and used the integral representation for the Bessel function $J_n(x)$ in Ref.~\cite{tem96}. Inserting Eq.~\eqref{a2_6} in Eq.~\eqref{a2_5} and using $J_{-n}(|\mathbf{K}|)=(-1)^{n}J_n(|\mathbf{K}|)$, we get the eigenvalues
\begin{eqnarray}
\lambda_j\!&=&\!QJ_0(|\mathbf{K}|) -C^{(1)}[e^{ij\theta}]_j\nonumber\\
\!&-&\!Q^2\sum_{l, l\neq j}\frac{(-1)^{l-j} [J_{l-j}(|\mathbf{K}|)]^2}{C^{(1)}[e^{ij\theta}]_j-C^{(1)}[e^{il\theta}]_l}, \label{a2_7}
\end{eqnarray}
up to terms of order $\varepsilon^3$. The last term on the right hand side of Eq.~\eqref{a2_7} is smaller than the other two and we can solve the equation $\lambda_j=0$ by iteration, thereby finding the approximate solution
\begin{eqnarray}
Q_j=\frac{C^{(1)}[e^{ij\theta}]_j}{J_0(|\mathbf{K}|)}\!\left[1 + \sum_{l, l\neq j}\frac{(-1)^{l-j} \!\left[\frac{J_{l-j}(|\mathbf{K}|)}{J_0(|\mathbf{K}|)}\right]^2}{1-\frac{C^{(1)}[e^{il\theta}]_l}{C^{(1)}[e^{ij\theta}]_j}}\right]\!. \label{a2_8}
\end{eqnarray}
For $j=1$, the largest term in this sum has index $l=2$ and is of order $|\mathbf{K}|^2$, whereas all other terms are $O(|\mathbf{K}|^4)$ and smaller as $|\mathbf{K}|\to 0$. Ignoring them, Eq.~\eqref{a2_8} yields 
\begin{eqnarray}
Q_1=\frac{C^{(1)}[e^{i\theta}]_1}{J_0(|\mathbf{K}|)}\!\left[1 - \frac{ \!\left[\frac{J_{1}(|\mathbf{K}|)}{J_0(|\mathbf{K}|)}\right]^2}{1-\frac{C^{(1)}[e^{i2\theta}]_2}{C^{(1)}[e^{i\theta}]_1}}\right]\!, \label{a2_9}
\end{eqnarray}
which is Eq.~\eqref{eq19}. For small $|\mathbf{K}|$, this equation is equivalent to 
\begin{eqnarray}
Q_1\approx\frac{C^{(1)}[e^{i\theta}]_1}{J_0(|\mathbf{K}|)+\frac{[J_1(|\mathbf{K}|)]^2C^{(1)}[e^{i\theta}]_1}{J_0(|\mathbf{K}|)\, (C^{(1)}[e^{i\theta}]_1-C^{(1)}[e^{i2\theta}]_2)}}, \label{a2_10}
\end{eqnarray}
which is Eq.~(26) of Ref.~\cite{BT18}.

\paragraph{Bifurcation theory using multiple scales.} For Hopf and period doubling bifurcations, the scaling of space and time is diffusive and we can use the multiple scales ansatz \cite{BT10,neu15}
\begin{eqnarray}
f(\theta,\textbf{x},t;\epsilon)=f_0+\sum_{j=1}^\infty\epsilon^j f^{(j)}(\theta,t,\textbf{X},T), \label{a2_11}
\end{eqnarray}
with $\textbf{X}=\epsilon\textbf{x}$, $T=\epsilon^2t$, $\eta=\eta_c+\epsilon^2\eta_2$, $\rho=\rho_0+\epsilon^2r(\mathbf{X})$. Inserting this ansatz into Eq.~\eqref{eq8}, we get a hierarchy of linear equations that have to be solved recursively:
\begin{eqnarray}
\mathcal{L}f^{(1)}\!\equiv f^{(1)}(\theta,t+1)- C^{(1)}[ f^{(1)}](\theta,t)=0,\label{a2_12}
\end{eqnarray}
\begin{eqnarray}
\mathcal{L}f^{(2)}=-\textbf{v}\!\cdot\!\nabla_X f^{(1)}(\theta,t+1) + C^{(2)}[f^{(1)},f^{(1)}](\theta,t),\quad\label{a2_13}
\end{eqnarray}
\begin{eqnarray}
&&\mathcal{L}f^{(3)}=-\left(\frac{\partial}{\partial T}+\frac{1}{2}(\textbf{v}\!\cdot\!\nabla_X)^2\right)\! f^{(1)}(\theta,t+1)
\nonumber\\
&&\quad\quad\,\,-\,\textbf{v}\!\cdot\!\nabla_X f^{(2)}(\theta,t+1)+\eta_2\frac{\partial C^{(1)}[f^{(1)}(\theta,t)]}{\partial\eta}\big|_{\eta_c}\nonumber\\
&&\quad\quad\,\, +\, r\,\frac{\partial C^{(1)}[f^{(1)}(\theta,t)]}{\partial\rho}\big|_{\eta_c}\!+2 C^{(2)}[f^{(1)},f^{(2)}](\theta,t)\nonumber\\
&&\quad\quad\,\, +\, C^{(3)}[f^{(1)},f^{(1)},f^{(1)}](\theta,t), \label{a2_14}
\end{eqnarray}
etc. In these equations, we have omitted that the $f^{(j)}$ also depend on $\textbf{X}$ and $T$. $C^{(2)}[f,g]$ and $C^{(3)}[f,g,h]$ are quadratic and cubic functionals, respectively, resulting from the expansion of $C_E[f_0+\epsilon\tilde{f}]$ in powers of $\epsilon$; cf. Eqs.~\eqref{eq32}-\eqref{eq33}. The solution of Eq.~\eqref{a2_12} is Eq.~\eqref{eq25} with $T=\epsilon^2t$. Eq.~\eqref{a2_13} has the solution \eqref{eq35}. Inserting Eqs.~\eqref{eq25} and \eqref{eq36} into Eq.~\eqref{a2_14} and using the solvability condition, we obtain the CGLE \eqref{eq41}. Its uniform solution with $r=0$ is Eq.~\eqref{eq42}.

\section{Coefficients of the amplitude equations}\label{ap:3}
\subsection{Hopf and period doubling bifurcations}
The coefficients $\delta$ and $\mu$ of the amplitude equation, Eq.~\eqref{eq38}, as obtained from the solvability condition for Eq.~\eqref{eq31}, are:
\begin{eqnarray}
&&\delta= \frac{1}{8}\!\left(\frac{1+e^{-i\Omega}C^{(1)}[e^{i2\theta}]_2}{1-e^{-i\Omega}C^{(1)}[e^{i2\theta}]_2}-i\cot\frac{\Omega}{2}\right)\!,\label{a3_1}\\
&&\mu=- \frac{2C^{(2)}[e^{i\theta}\!,e^{i\theta}]_2C^{(2)}[e^{i2\theta}\!,e^{-i\theta}]_1\!}{e^{i\Omega}\{e^{i2\Omega}\!-\!C^{(1)}[e^{i2\theta}]_2\}}\nonumber\\
&&- \frac{C^{(3)}[e^{i\theta}\!,e^{i\theta}\!,e^{-i\theta}]_1}{e^{i\Omega}}\!\sim \frac{\pi^4R_0^4/M}{1-e^{-i2\Omega}C^{(1)}[e^{i2\theta}]_2}\nonumber\\
&&=\frac{\pi^3R_0^2/\rho_0}{1-e^{-i2\Omega}C^{(1)}[e^{i2\theta}]_2}.   \label{a3_2}
\end{eqnarray}
Here $C^{(1)}[e^{i2\theta}]_2=Q_2$ is given by Eq.~\eqref{eq21}, and
\begin{widetext}
\begin{eqnarray}
&&C^{(2)}[e^{i\theta}\!,e^{i\theta}]_2=8\pi^2 R_0^2e^{-M}
\sum_{n=2}^N\frac{n M^{n-2}}{(n-2)!}K^{11}_{cc}(n)\int_{-\pi}^\pi e^{-i2\xi}g(\xi)d\xi\sim\pi^2R_0^2 \int_{-\pi}^\pi e^{-i2\xi}g(\xi)d\xi,\label{a3_3}\\
&&C^{(2)}[e^{i2\theta}\!,e^{-i\theta}]_1\!=\!4\pi^2R_0^2e^{-M}
\sum_{n=2}^N\frac{nM^{n-2}}{(n-2)!}K^{1}_{c2c}(n)\int_{-\pi}^\pi e^{-i\xi}g(\xi)d\xi\sim-\frac{\pi^{5/2}R_0^2}{8\sqrt{M}}\int_{-\pi}^\pi e^{-i\xi}g(\xi)d\xi,\label{a3_4}\\
&& C^{(3)}[e^{i\theta}\!,e^{i\theta}\!,e^{-i\theta}]_1\!=\!\frac{16}{3}\pi^4R_0^4e^{-M}
\sum_{n=3}^N\frac{nM^{n-3}}{(n-3)!}K^{1}_{ccc}(n)\int_{-\pi}^\pi e^{-i\xi}g(\xi)d\xi\sim-\frac{\pi^{9/2}R_0^4}{2\sqrt{M}}\int_{-\pi}^\pi e^{-i\xi}g(\xi)d\xi,   \label{a3_5}
\end{eqnarray}
\end{widetext}
in the limit as $M\gg 1$ (cf.\ Section 5 of Ref.~\cite{BT18}). In this limit, $|C^{(1)}[e^{i2\theta}]_2|=|\int_{-\pi}^\pi e^{-i2\xi}g(\xi)d\xi|/2\leq 1/2$, and therefore
\begin{eqnarray}
\mbox{Re }\delta=\frac{1-|C^{(1)}[e^{i2\theta}]_2|^2}{8|1-e^{-i\Omega}C^{(1)}[e^{i2\theta}]_2|^2}\Longrightarrow\nonumber\\
\mbox{Re }\delta\geq \frac{3}{32}\frac{1}{(1+|C^{(1)}[e^{i2\theta}]_2|)^2}>\frac{1}{24}>0. \label{a3_6}
\end{eqnarray}
Similarly, 
\begin{eqnarray}
\mbox{Re }\mu=\frac{\pi^4R_0^4}{M}\frac{1-\mbox{Re}(e^{-i2\Omega}C^{(1)}[e^{i2\theta}]_2)}{|1-e^{-2i\Omega}C^{(1)}[e^{i2\theta}]_2|^2}\Longrightarrow\nonumber\\
\mbox{Re }\mu\geq \frac{\pi^4R_0^4}{M}\frac{1-|C^{(1)}[e^{i2\theta}]_2|}{(1+|C^{(1)}[e^{i2\theta}]_2|)^2}>\frac{2\pi^4R_0^4}{9M}>0.\label{a3_7}
\end{eqnarray}

\subsection{Pitchfork bifurcation}
In this case, all the coefficients in the amplitude equations are real. The coefficients $\delta$ and $\mu$ are given by the real part of Eq.~\eqref{a3_1} with $\Omega=0$ and by Eq.~\eqref{a3_2} with $\Omega=0$, respectively. Thus, both $\delta$ and $\mu$ are positive in the limit as $M\gg 1$. The other coefficients appearing in Eqs.~\eqref{eq55}-\eqref{eq56} are
\begin{eqnarray}
 Q_\eta\!=\!\frac{\partial}{\partial\eta}\!\left(\ln C^{(1)}[e^{i\theta}]_1\right)\big|_{\eta_c}\!, \,\delta=\frac{1+C^{(1)}[e^{i2\theta}]_2}{8(1-C^{(1)}[e^{i2\theta}]_2)},\nonumber\\
\gamma_1=\frac{1}{4}C^{(2)}[1,e^{i\theta}]_1-\frac{C^{(2)}[e^{i\theta},e^{i\theta}]_2}{2(1-C^{(1)}[e^{i2\theta}]_2)},\nonumber\\
\gamma_2=-\frac{C^{(2)}[e^{-i\theta},e^{i2\theta}]_1}{1-C^{(1)}[e^{i2\theta}]_2},\,\,\gamma_3=\frac{C^{(2)}[1,e^{i\theta}]_1}{2}, \label{a3_8}
\end{eqnarray}
In the limit as $M\gg 1$, we can use Eqs.~\eqref{a3_3}-\eqref{a3_5} together with
\begin{eqnarray}
C^{(2)}[1,e^{i\theta}]_1&=&\frac{\pi M}{\rho_0}\frac{\partial C^{(1)}[e^{i\theta}]_1}{\partial M}\nonumber\\
&\sim&\frac{\pi C^{(1)}[e^{i\theta}]_1}{2\rho_0}=\frac{\pi}{2\rho_0},\nonumber\\
C^{(3)}[1,1,e^{i\theta}]_1\!\!&=&\!\frac{2\pi^4\! R_0^4}{3}\frac{\partial^2C^{(1)}[e^{i\theta}]_1}{\partial M^2}\nonumber\\
&\!\sim\!&-\frac{\pi^2C^{(1)}[e^{i\theta}]_1}{6\rho_0^2}=-\frac{\pi^2}{6\rho_0^2},\label{a3_9}
\end{eqnarray}
to calculate these coefficients. Recall that $C^{(1)}[e^{i\theta}]_1=1$ at $\eta_c$. We obtain:
\begin{eqnarray}
&&\frac{C^{(3)}[1,1,e^{i\theta}]_1\!-\!2 (C^{(2)}[1,e^{i•\theta}]_1)^2}{4\pi^2}\!\sim-\frac{1}{6\rho_0^2},\nonumber\\
&&Q_\eta\sim - \frac{\sqrt{\pi M}}{2\eta_c}\!\left( \frac{2}{\sqrt{\pi M}}- \cos\frac{\eta_c}{2}\right)\!,\nonumber\\
&&\gamma_1\sim\pi^2R_0^2\!\left(1+\frac{1}{8M}-\gamma_{0}\right)\!,\quad\gamma_2\sim\frac{\gamma_0\pi^2R_0^2}{4M}, \nonumber\\
&&\gamma_0=\frac{1}{1-\frac{1}{\sqrt{\pi M}}\cos\frac{\eta_c}{2}-\frac{p}{2}\left(1+\cos\frac{\eta_c}{2}\right)},\nonumber\\
&&\gamma_3\sim\frac{\pi^2R_0^2}{4M},\quad\delta\sim\frac{2\gamma_0-1}{8}, \quad\mu\sim\frac{\pi^4R_0^4\gamma_0}{M}. \label{a3_10}
\end{eqnarray}
To find $\gamma_0$ in the limit as $M\gg 1$, we have used its definition, $\gamma_0^{-1}=1-C^{(1)}[e^{i2\theta}]_2$, and the linear stability condition, $Q_1=1$ (for $\xi_0=\pi$):
\begin{eqnarray}
&&-p+2\frac{1-p}{\eta_c}\sin\frac{\eta_c}{2}= \frac{2}{\sqrt{\pi M}}\nonumber\\
&&\Longrightarrow\frac{1-p}{\eta_c}\sin\frac{\eta_c}{2}=\frac{p}{2}+\frac{1}{\sqrt{\pi M}},\label{a3_11}
\end{eqnarray}
which, inserted in Eq.~\eqref{eq21} for $C^{(1)}[e^{i2\theta}]_2$, yields
\begin{eqnarray}\nonumber
&&C^{(1)}[e^{i2\theta}]_2\!=\frac{1}{2}\!\left(p+\frac{1-p}{\eta_c}\sin\eta_c\right)\!=\frac{p}{2}\\
&&+\frac{1-p}{\eta_c}\sin\frac{\eta_c}{2}\cos\frac{\eta_c}{2}=\frac{p}{2}\!\left(1+\cos\frac{\eta_c}{2}\right)\!\nonumber\\
&&+ \frac{1}{\sqrt{\pi M}}\cos\frac{\eta_c}{2}.\label{a3_12}
\end{eqnarray}

\end{document}